\documentclass[12pt]{iopart}

\usepackage{iopams,subeqn} 

\usepackage{graphicx}

\newcommand{\eqref}{\eref}

\newcommand{\Jcal}{\mathcal{J}}

\begin{document}

\title{Interaction vs inhomogeneity in a periodic TASEP}

\author{Beatrice Mina$^{1}$, Alex Paninforni$^{2}$, Alessandro Pelizzola$^{2,3,4}$ and Marco Pretti$^{4}$}

\address{$^1$ Istituto Universitario di Studi Superiori (IUSS) - Palazzo del Broletto, Piazza della Vittoria 15, 27100 Pavia, Italy}
\address{$^2$ Dipartimento Scienza Applicata e Tecnologia (DISAT), Politecnico di Torino - Corso Duca degli Abruzzi 24, 10129 Torino, Italy}
\address{$^3$ INFN, Sezione di Torino - Via Pietro Giuria 1, 10125 Torino, Italy}
\address{$^4$ Consiglio Nazionale delle Ricerche - Istituto Sistemi Complessi (CNR-ISC) c/o DISAT, Politecnico di Torino - Corso Duca degli Abruzzi 24, 10129 Torino, Italy}

\ead{marco.pretti@polito.it}

\begin{abstract}
	We study the non-equilibrium steady states in a totally-asymmetric simple-exclusion process with periodic boundary conditions, also incorporating (i) an extra (nearest-neighbour) repulsive interaction and (ii) hopping rates characterized by a smooth spatial inhomogeneity. 
	We make use of a generalized mean-field approach (at the level of nearest-neighbour pair clusters), in combination with kinetic Monte Carlo simulations. 
	It turns out that the so-called shock phase can exhibit a lot of qualitatively different subphases, including multiple-shock phases, and a minimal-current shock phase. 
	We argue that the resulting, considerably rich phase diagram should be relatively insensitive to minor details of either interaction or spatial inhomogeneity. 
	As a consequence, we also expect that our results help elucidate the nature of shock subphases detected in previous studies. 
\end{abstract}



\submitto{J. Phys. A: Math. Theor.}

\maketitle

\section{Introduction}

The \emph{totally-asymmetric simple-exclusion process} (TASEP) is an elementary stochastic-transport model, with a variety of applications to out-of-equilibrium physics, ranging from biological microsystems to vehicular traffic~\cite{ChouMallickZia2011,SchadschneiderChowdhuryNishinari2011}, and intriguing theoretical connections (Kardar-Parisi-Zhang universality, random matrix theory)~\cite{KriecherbauerKrug2010,Lazarescu2015}. 
For such reasons, and since it is simple enough that it can be treated analytically up to a certain extent, this model has inspired a lot of fundamental studies in non-equilibrium statistical mechanics. 
The model is usually defined on a linear chain, whose nodes can be occupied by at most one particle, and where each particle can hop to the adjacent node, provided the latter is empty: \emph{totally asymmetric} means that hopping can occur in one direction only. 
Over the years, many variants of the model have been investigated, with the aim of incorporating extra features, which may be relevant in different physical contexts. 
For example, a first important issue is the distinction between periodic or open boundary conditions. 
In the latter case, if the chain is coupled to particle reservoirs that inject and extract particles at opposite ends, the system exhibits different non-equilibrium steady states and a number of phase transitions among them, controlled by injection and extraction rates (boundary-induced phase transitions)~\cite{Krug1991}. 
This noticeable case is still exactly solvable, and the steady-state solution dates back to the 1990s~\cite{DerridaDomanyMukamel1992,SchutzDomany1993,DerridaEvansHakimPasquier1993,Derrida1998}. 

Among other interesting issues, that of spatial inhomogeneity of hopping rates has been frequently considered in the literature. 
In case of periodic boundary conditions, it turns out that several different types of inhomogeneities (a unique slower rate \cite{JanowskyLebowitz1992,JanowskyLebowitz1994,Kolomeisky1998}, randomly distributed rates \cite{TripathyBarma1998,BengrineBenyoussefEzZahraouyMhirech1999}, smoothly varying rates \cite{StinchcombeDeQueiroz2011,BanerjeeBasu2020,GoswamiChatterjeeMukherjee2022}) give rise to the same kind of phenomena, namely, steady states featuring domain walls (or \emph{shocks}), which separate regions at different densities. 
Similar shocks can also emerge if the system is allowed to adsorb and/or desorb particles in its bulk (Langmuir kinetics) with proper size scaling \cite{PopkovRakosWillmannKolomeiskySchutz2003,EvansJuhaszSanten2003,ParmeggianiFranoschFrey2003,ParmeggianiFranoschFrey2004,BottoPelizzolaPrettiZamparo2019,BottoPelizzolaPrettiZamparo2020}. 
Note that, in such cases, the rates are spatially uniform, but the density profile is not (even in the steady state), due to violation of the particle-number conservation constraint. 
A different type of extra feature, that can be incorporated in the model, is some kind of interaction beyond the basic exclusion constraint, in the simplest case a nearest-neighbour (NN) interaction \cite{KatzLebowitzSpohn1984,PopkovSchutz1999,HagerKrugPopkovSchutz2001,DierlMaassEinax2011,DierlMaassEinax2012,DierlEinaxMaass2013,MidhaKolomeiskyGupta2018jstat,PelizzolaPrettiPuccioni2019,AntalSchutz2000,BottoPelizzolaPretti2018,BaumgaertnerNarasimhan2023}. 
Such interactions can give rise to many different non-equilibrium steady states, and therefore very complex phase diagrams (in particular, up to 7 phases for strong enough repulsive interaction and open boundary conditions) \cite{PopkovSchutz1999,HagerKrugPopkovSchutz2001,DierlMaassEinax2011,DierlMaassEinax2012,DierlEinaxMaass2013,MidhaKolomeiskyGupta2018jstat,PelizzolaPrettiPuccioni2019}. 

\begin{figure}
	\centerline{\includegraphics*[width=0.70\textwidth,bb=0 490 500 730]{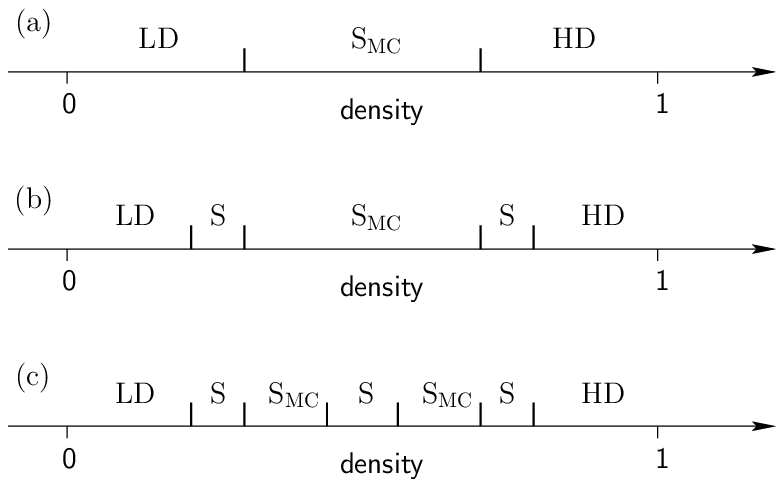}}
	\caption
	{
		Three regimes observed in \cite{PalGupta2021}, as a function of the (mean) particle density: (a) no interaction or weak interaction; (b) mildly strong interaction; (c) very strong interaction. 
		Phase tags are explained in the text. 
	}
	\label{fig:schema}
\end{figure}

Of course, it may also be of interest to investigate models incorporating more than one of the above features, as the latter may likely occur together in physical systems. 
For instance, \cite{PierobonMobiliaKouyosFrey2006} investigates a TASEP-like model with both hopping-rate inhomogeneity and Langmuir kinetics, which the authors argue to be minimal ``ingredients'' for modeling motor-protein motion along microtubules in living cells. 
Still motivated by biological transport phenomena, as well as by methodological issues, \cite{MidhaKolomeiskyGupta2018pre} investigates the role of NN interactions in a TASEP with Langmuir kinetics, though without inhomogeneities. 
Furthermore, \cite{JindalMidhaGupta2020,PalGupta2021} investigate the interplay between NN interactions and rate inhomogeneities, in the presence of open \cite{JindalMidhaGupta2020} or periodic \cite{PalGupta2021} boundary conditions. 
In particular, the recent work by Pal and Gupta \cite{PalGupta2021} (periodic boundary conditions) points out that a NN \emph{repulsive} interaction, in combination with a very simple rate inhomogeneity (represented by a regular function with a single minimum and a single maximum) gives rise to a seemingly non-trivial phenomenology.
The presence of three different regimes is revealed, as the repulsion strength increases (see figure~\ref{fig:schema}). 
In case of weak repulsion (figure~\ref{fig:schema}a), the behavior of the system is qualitatively equivalent to that of a system without NN interaction \cite{BanerjeeBasu2020}, with a single phase at intermediate densities, characterized by the presence of a shock in the density profile and maximal current (S${}_\mathrm{MC}$ phase), and by two phases, at high and low density, characterized by a spatially-modulated density profile but without shock, and less-than-maximal current (HD and LD phases).
This regime corresponds to a (local) current-density relation featuring a unique ``central maximum'', at density 1/2.
In case of stronger repulsion, where it is known that the current-density relation develops a ``central minimum'' at density 1/2 and two ``side maxima'', the other two regimes emerge. 
In one of them, at ``mildly strong'' interaction (figure~\ref{fig:schema}b), the S${}_\mathrm{MC}$ phase is characterized by two shocks, one of which (the ``descending'' one) localized at the minimum-rate position.
Furthermore, the separation regions, between this phase and the HD and LD phases, are characterized by a single shock, somewhat similar to the ``ordinary'' shock phase (i.e.~that of a model without NN interaction), but with apparently non-maximal current.
The nature of these phases (called ``S phases'') remains not completely clarified, since in this regime there appear large discrepancies between theory and simulations, and the analysis relies mainly on the latter, which are unfortunately affected by significant finite-size effects.
In the other regime, at ``very strong'' interaction (figure~\ref{fig:schema}c), the presence of yet another S phase is observed at densities close to 1/2, with a minimum of the current between the two maximal-current S${}_\mathrm{MC}$ regions.

In this paper we aim at elucidating the phenomenology that emerges from the work described above, which we find quite interesting, especially since, as previously discussed, it originates from the interplay of just two simple extra ``ingredients'' of the model, compared to ordinary TASEP (rate inhomogeneity and repulsive NN interaction). 
To do this, we are going to study a model in principle identical to that of Pal and Gupta, but with a different constraint on the rates, which we will call the KLS condition (since it was originally introduced by Katz, Lebowitz and Spohn \cite{KatzLebowitzSpohn1984,PopkovSchutz1999,HagerKrugPopkovSchutz2001}). 
The underlying idea is explained below. 
On the one hand, we expect that the overall physical behaviour remains qualitatively unaffected by the change in the rate constraint, specifically because it can be seen that the modified model undergoes the same kind of ``transition'' (from unimodal to bimodal) in the current-density relation. 
On the other hand, it is known that the steady state of the model with KLS condition, in the simpler case of homogeneous rates, coincides with the equilibrium state of a one-dimensional lattice gas with NN interaction (a detailed proof is given by Dierl, Einax and Maass~\cite{DierlEinaxMaass2013}). 
As a consequence, a generalized mean-field theory including NN pair correlations, which we denote as \emph{pair approximation} (PA), is exact in this case. 
Thus, for our system, characterized by the assumption of smoothly-varying rates, we expect that the PA remains very accurate. 
Indeed, all the numerical evidences collected in this work suggest that it exactly reproduces the smooth sections of the density profiles, so that we can obtain a very detailed analysis of the various phases, practically without finite-size effects.

The paper is organized as follows. 
In sections \ref{sec:model} and \ref{sec:methods} we introduce the model and the PA theory, respectively. 
Section \ref{sec:density_profiles} describes, in terms of density profiles, the different regimes occurring in the non-equilibrium steady states of the model. 
The results are made more systematic in section \ref{sec:phase_diagrams}, which presents and explains the analytical phase diagrams, and the way they have been worked out. 
Section \ref{sec:numerical_simulations} is devoted to numerical simulations, in order to check the accuracy of the PA. 
We recap our findings and draw some conclusions in section~\ref{sec:conclusions}.
The technical details are reported in four appendices.

\section{The model}
\label{sec:model}

The model we consider is defined on a linear chain of $L$ nodes, labelled as ${i = 0,\dots,L-1}$, with periodic boundary conditions. 
Each node can be empty or occupied by a single particle. 
We introduce occupation-number variables $n_i^t$, taking value $0$ or $1$ if node $i$ at time $t$ is respectively empty or occupied. 
Each particle can hop in only one direction (conventionally from node $i$ to node ${i+1}$), provided the destination node is empty. 
The rate of hopping from node $i$ (to ${i+1}$) may depend on both the position $i$ and on the ${i-1}$ and ${i+2}$ node configurations, so that it will be denoted as ${w}_{i}(n_{i-1},n_{i+2})$. 
This type of dependence can be regarded as a NN interaction. 
Node indices are always understood modulo $L$, according to periodic boundary conditions. 

To begin with, we shall consider the most general case where, for each node $i$, we have 4 different hopping rates (associated with all possible occupancy states of the 2 nodes ${i-1}$ and ${i+2}$), with a fully arbitrary dependence on the position $i$. 
The PA theory will be first developed in such a general case. 
Subsequently, we shall take some more restrictive hypotheses, along the lines of \cite{PalGupta2021} and previous literature~\cite{DierlMaassEinax2011,DierlMaassEinax2012,DierlEinaxMaass2013,MidhaKolomeiskyGupta2018jstat,MidhaKolomeiskyGupta2018pre,JindalMidhaGupta2020}. 
In particular, we shall assume the following expressions for the rates, whose meaning is discussed below: 
\begin{subequations} 
\label{eq:gamma}
\begin{eqnarray}
	{w}_{i}(0,0) & \triangleq {p} \, \lambda(i/L) 
	\, , \label{eq:gamma00} \\
	{w}_{i}(0,1) & \triangleq {q} \, \lambda(i/L) 
	\, , \label{eq:gamma01} \\
	{w}_{i}(1,0) & \triangleq {r} \, \lambda(i/L) 
	\, , \label{eq:gamma10} \\
	{w}_{i}(1,1) & \triangleq {s} \, \lambda(i/L) 
	\, . \label{eq:gamma11} 
\end{eqnarray}
\end{subequations} 
The first restrictive assumption contained in \eqref{eq:gamma} is that the spatial dependence of the rates is fully incorporated in a unique function $\lambda(x)$, being independent of the occupancy states of the forward and backward nodes. 
As a consequence, the \emph{ratios} between rates associated to different forward/backward occupancies are specified only by the position-independent coefficients ${p}$, ${q}$, ${r}$, ${s}$. 
In the following we will often call $\lambda(x)$ the \emph{rate modulation function}. 
The second important assumption is the \emph{continuity} of the rate modulation function, meaning in practice that the spatial dependence of the rates is ``smooth'', as it occurs only through the \emph{scaled} position variable ${x \triangleq i/L}$ (in order to respect boundary conditions, we can assume that $\lambda(x)$ is periodic with period~$1$). 
Furthermore, as anticipated in the introduction, we shall assume that the rate modulation function is characterized by a unique minimum and a unique maximum (over its period). 
In particular we shall consider a sinusoidal function, taking values in a positive interval $[\lambda_{\min},\lambda_{\max}]$, that is
\begin{equation}
	\lambda(x) \triangleq \frac{\lambda_{\max} + \lambda_{\min}}{2} + \frac{\lambda_{\max} - \lambda_{\min}}{2} \cos(2 \pi x)
	\, .
	\label{eq:rmf}
\end{equation} 
With this particular choice (irrelevant to the qualitative behaviour of the model) the rate modulation function turns out to be completely specified by the only 2 parameters $\lambda_{\min}$ and $\lambda_{\max}$. 
Throughout our analysis, we shall usually consider one of them (for instance $\lambda_{\min}$) fixed as a reference value, varying the ratio $\lambda_{\max}/\lambda_{\min}$, which we shall call \emph{rate modulation ratio}.

We shall also introduce some restrictive hypotheses on the ${p},{q},{r},{s}$ parameters. 
As in Pal and Gupta's work~\cite{PalGupta2021}, we shall first assume ${{p}={s}}$, which turns out to induce a particle-hole symmetry.\footnote{Throughout the paper, we shall often use the term ``particle-hole symmetry'' to denote the mirror symmetry of the current-density relation, although this is just one aspect of the symmetry.} 
In fact, one of the four parameters can be set equal to $1$ without loss of generality, so the imposed condition will actually be
\begin{equation}
	{p} = {s} = 1
	\, ,
	\label{eq:simmetria_parametri}
\end{equation}
leaving ${q}$ and ${r}$ as free parameters. 
Actually, we find it more convenient to define 
\begin{equation}
	{v} \triangleq \sqrt{\frac{\,{q}\,}{{r}}} 
	\label{eq:definizione_v}
	\, ,
\end{equation} 
and thence to consider ${v}$ and ${q}$ as independent parameters, with ${r}$ fixed by \eqref{eq:definizione_v}. 
We shall call ${v}$ the \emph{interaction parameter}, as it discriminates the attractive case (${{q}>{r}}$, or ${{v}>1}$) from the repulsive one (${{q}<{r} }$, or ${{v}<1}$). 
It can be seen that the situation studied in \cite{PalGupta2021} corresponds to adding the simple condition
\begin{equation}
	{q} \, {r} = {p} \, {s} 
	\label{eq:condizione_DME}
	\, ,
\end{equation} 
which, together with the symmetry hypothesis \eqref{eq:simmetria_parametri} and definition \eqref{eq:definizione_v}, leads to 
\begin{subequations} 
\begin{eqnarray}
	{q} = {v} 
	\label{eq:condizione_DME_vq}
	\, , \\
	{r} = 1/{v} 
	\label{eq:condizione_DME_vr}
	\, . 
\end{eqnarray} 
\end{subequations} 
This assumption is based on a thermodynamic consistency argument, and was previously proposed by Dierl, Maass and Einax \cite{DierlMaassEinax2011}, so we shall denote it as the DME condition. 
Conversely, as anticipated in the introduction, we will assume the KLS condition, namely
\begin{equation}
	{q} + {r} = {p} + {s} 
	\, ,
	\label{eq:condizione_KLS}
\end{equation}
which, still in combination with \eqref{eq:simmetria_parametri} and \eqref{eq:definizione_v}, entails 
\begin{subequations} 
\label{eq:condizioni_KLS}
\begin{eqnarray}
	{q} = \frac{2{v}^2}{1+{v}^2}
	\label{eq:condizione_KLS_vq}
	\, , \\
	{r} = \frac{2}{1+{v}^2} 
	\label{eq:condizione_KLS_vr}
	\, . 
\end{eqnarray}
\end{subequations} 
It is worth noting that, even though all the physical results that we report in the article have been obtained \emph{with} the KLS condition, most of the related analytical theory is developed without it, in particular retaining only the symmetry assumption \eqref{eq:simmetria_parametri} and keeping both ${v}$ and ${q}$ as free parameters.

\section{The analytical methods}
\label{sec:methods}

Let us now present the PA, in the most general case introduced above. 
It is useful to define first some notation for marginal distributions and expectation values. 
Let us denote the marginal distribution at time $t$, for a cluster of consecutive nodes starting at $i$, by 
\begin{equation}
	{P}_{i}^{t}[klm\dots] \triangleq \mathbb{P} \left\{ 
	n_{i  }^{t} = k \, , 
	n_{i+1}^{t} = l \, , 
	n_{i+2}^{t} = m \, , 
	\dots \right\}
	\, .
	\label{eq:marginal}
\end{equation}
Moreover, let us define specific symbols for the local densities and NN correlations, respectively: 
\begin{eqnarray}
	\rho_{i}^{t}  & \triangleq \langle n_{i}^{t} \rangle 
	\, , \label{eq:rhoit} \\
	\phi_{i}^{t}  & \triangleq \langle n_{i}^{t} n_{i+1}^{t} \rangle 
	\, . \label{eq:phiit}
\end{eqnarray}
As we are dealing with binary random variables, the latter quantities completely specify 1-node and 2-node marginals (for NN pairs), and can be used to parameterize them as follows: 
\begin{subequations} 
\label{eq:site_marginal}
\begin{eqnarray}
	{P}_{i}^{t}[1]  & = \rho_{i}^{t}
	\, , \\
	{P}_{i}^{t}[0]  & = 1 - \rho_{i}^{t}
	\, , 
\end{eqnarray}
\end{subequations} 
and 
\begin{subequations} 
\label{eq:pair_marginal}
\begin{eqnarray}
	{P}_{i}^{t}[11] & = \phi_{i}^{t}
	\, , \\
	{P}_{i}^{t}[10] & = \rho_{i}^{t}   - \phi_{i}^{t}
	\, , \\
	{P}_{i}^{t}[01] & = \rho_{i+1}^{t} - \phi_{i}^{t}
	\, , \\
	{P}_{i}^{t}[00] & = 1 - \rho_{i}^{t} - \rho_{i+1}^{t} + \phi_{i}^{t}
	\, . 
\end{eqnarray}
\end{subequations} 
Now, from the master equation one can derive time-evolution equations for $\rho_{i}^{t}$ and $\phi_{i}^{t}$, through a marginalization procedure (a detailed derivation is given in \cite{PelizzolaPrettiPuccioni2019}). 
As far as the local densities are concerned, we obtain a typical continuity equation, namely
\begin{eqnarray}
	\dot{\rho}_{i}^{t} = \Jcal_{i-1}^{t} - \Jcal_{i}^{t}
	\, , 
	\label{eq:continuity}
\end{eqnarray}
where $\Jcal_{i}^{t}$ represents the probability current from $i$ to ${i+1}$ at time $t$. 
Such current can be written as a sum of 4 contributions, one for each possible combination of backward and forward occupation states, in formulae 
\begin{equation}
	\Jcal_{i}^{t} 
	= \Jcal_{i}^{t}(0,0) + \Jcal_{i}^{t}(0,1) + \Jcal_{i}^{t}(1,0) + \Jcal_{i}^{t}(1,1)
	\, , 
	\label{eq:corrente}
\end{equation}
where 
\begin{equation}
	\Jcal_{i}^{t}(k,n) = {w}_{i}(k,n) \, {P}_{i-1}^{t}[k10n] 
	\, . 
	\label{eq:corrente_elem}
\end{equation}
Regarding local correlations, we have the following equation
\begin{eqnarray}
	\dot{\phi}_{i}^{t} 
	= \Jcal_{i-1}^{t}(0,1) + \Jcal_{i-1}^{t}(1,1) - \Jcal_{i+1}^{t}(1,0) - \Jcal_{i+1}^{t}(1,1)
	\, . 
	\label{eq:phipunto}
\end{eqnarray}
In the end we can see that the time-derivatives of $\rho_{i}^{t}$ and $\phi_{i}^{t}$ can be written in terms of 4-node marginals, so the resulting time-evolution equations, though exact, are not closed. 
A possible closure scheme is naturally provided by the PA \cite{Pelizzola2005}, also known as Bethe approximation \cite{PlischkeBergersen1994} or 2-cluster approximation \cite{SchadschneiderChowdhuryNishinari2011}.
In the specific case, such approximation reads 
\begin{equation}
	{P}_{i}^{t}[k10n] \cong \frac
	{{P}_{i}^{t}[k1] \, {P}_{i+1}^{t}[10] \, {P}_{i+2}^{t}[0n]}
	{{P}_{i+1}^{t}[1] \, {P}_{i+2}^{t}[0]} 
	\, ,
	\label{eq:pairwapp}
\end{equation}
where 1-node (site) and 2-node (NN pair) marginals can be expressed in terms of $\rho_{i}^{t}$ and $\phi_{i}^{t}$ through equations \eqref{eq:site_marginal} and \eqref{eq:pair_marginal}. 
An analogous technique has already been applied to the NN-interacting TASEP, for instance in \cite{MidhaKolomeiskyGupta2018pre} (for a system with Langmuir kinetics) and \cite{JindalMidhaGupta2020} (with site-dependent hopping rates). 
In these papers, the resulting equations have been used as an intermediate step to determine a continuum limit, leading to slightly different methods, respectively denoted as \emph{cluster mean-field} and \emph{correlated cluster mean-field}. 
Conversely, in this paper we adopt a simpler strategy, as done in \cite{PelizzolaPrettiPuccioni2019}, that is, we perform a direct time-integration of the discrete system (at finite $L$), obtaining the steady state as a long-time limit of the numerical procedure (more precisely, we define the limit by requiring that the magnitude of all time derivatives stays below a certain threshold). 
This procedure does not need a considerable computational power, so we can easily reach quite large sizes (all the results reported in this section refer to ${L=10000}$), such that the density profiles obtained are (in most cases) practically indistinguishable from the continuum limit. 
Apart from this difference, our approach is equivalent to the cluster mean-field theory of \cite{PalGupta2021}, since both are based on the same pairwise factorization \eqref{eq:pairwapp}. 
We nonetheless prefer to retain the old-fashioned term PA, in order to avoid confusion with analogous approximation strategies, that take into account higher-order clusters~\cite{SchadschneiderChowdhuryNishinari2011,PelizzolaPretti2017}. 

In this work we shall also make use of the current-density relation, in a form derived from the PA theory.
This relation, which by definition refers to the continuum limit, is precisely a function returning the value of the current given that of the local density.
Such a function is completely specified by the local values of the hopping rates, which play the role of parameters. 
Collectively denoting by $\underline{w}$ an array of 4 possible rates (associated with the different occupancy states, as described above), we can write the current-density relation as 
\begin{equation}
	\Jcal = {F}_{\underline{w}}(\rho)
	\, .
	\label{eq:relazione_fondamentale}
\end{equation}
Since the rates $\underline{w}$ depend on the position, equation \eqref{eq:relazione_fondamentale} establishes a constraint to the spatial variations of the density $\rho$, such as to impose that the value of the current $\Jcal$ remains fixed.
Now, let us observe that, if all the hopping rates are multiplied by the same factor, say $\lambda$, the current gets multiplied by the same factor, in formulae
\begin{equation}
	{F}_{\lambda \underline{w}}(\rho)
	= \lambda {F}_{\underline{w}}(\rho)
	\, .
	\label{eq:omogeneita}
\end{equation}
In mathematical terms we may say that the function ${F}$ is homogeneous (of degree 1) with respect to the parameters.
This property follows from a simple physical argument, but it can also be verified explicitly (see \ref{app:cont_lim}). 
In our case it is specially important, since, as introduced above, the position dependence of the rates is given by a common prefactor, namely the rate modulation function $\lambda(x)$. 
As a consequence, in order to determine the possible density profiles, we need to invert a single current-density function (i.e.~the one characterized by the position-independent parameters ${p},{q},{r},{s}$), according to equation
\begin{equation}
	\rho(x) = {F}_{{p},{q},{r},{s}}^{-1} \left( \frac{\Jcal}{\lambda(x)} \right)
	\, .
	\label{eq:profili_analitici}
\end{equation}
For the model considered here, the inversion can be done analytically (see \ref{app:dens_prof}). 
In the following, we shall usually call \emph{reduced current} the (position-dependent) quantity defined as 
\begin{equation}
	{J}(x) \triangleq \frac{\Jcal}{\lambda(x)}
	\, .
	\label{eq:reduced_current}
\end{equation}
As known, the current-density function is not invertible in the narrow sense or, in other words, its inverse is a multi-valued function.
Therefore, equation \eqref{eq:profili_analitici} does not determine a unique density profile, but a (small) set of possible profiles.
The actual profile can be determined by imposing that the integral average of the function $\rho(x)$ is equal to the actual mean particle density in the system, say $\bar{\rho}$, which can thus be regarded as a control parameter. 
In formulae
\begin{equation}
	\bar{\rho} = \int_{0}^{1} \rho(x) \, \rmd x
	\, .
	\label{eq:densita_media}
\end{equation}
Since, according to equation \eqref{eq:profili_analitici}, the profile $\rho(x)$ depends not only on the parameters but also on the value of the current $\Jcal$, equation \eqref{eq:densita_media} determines the latter as a function of $\bar{\rho}$. 
It can be seen that, in certain ranges of $\bar{\rho}$ values, there does not exist a single continuous profile, being solution of \eqref{eq:profili_analitici}.
In such a case, the actual stationary profile is made up of continuous sections, each one being solution of \eqref{eq:profili_analitici}, separated by discontinuities (shocks).
Let us observe that, even in case of a single shock, equation \eqref{eq:densita_media} alone is unable to determine both resulting unknowns (namely current and shock position) simultaneously. 
However, it can be seen that in this situation the system dynamics determines the value of the current according to an extremal principle (maximum or possibly minimum), analogous to the one stated in \cite{PopkovSchutz1999,HagerKrugPopkovSchutz2001}.
Even with this extra condition, the shock position may not be completely determined (more so when there are more than one shocks), and in this case some shock-stability criterion must come into play.
We will also observe that, when a particular relation among the parameters occurs, the shock positions remain anyway indeterminate, and that this physically corresponds to the onset of two delocalized (but synchronized) shocks, in analogy to what has been reported by Banerjee and Basu (in a model with rate modulation function characterized by two equivalent absolute minima)~\cite{BanerjeeBasu2020}.

\begin{figure}
	\flushright{\includegraphics*[width=0.85\textwidth]{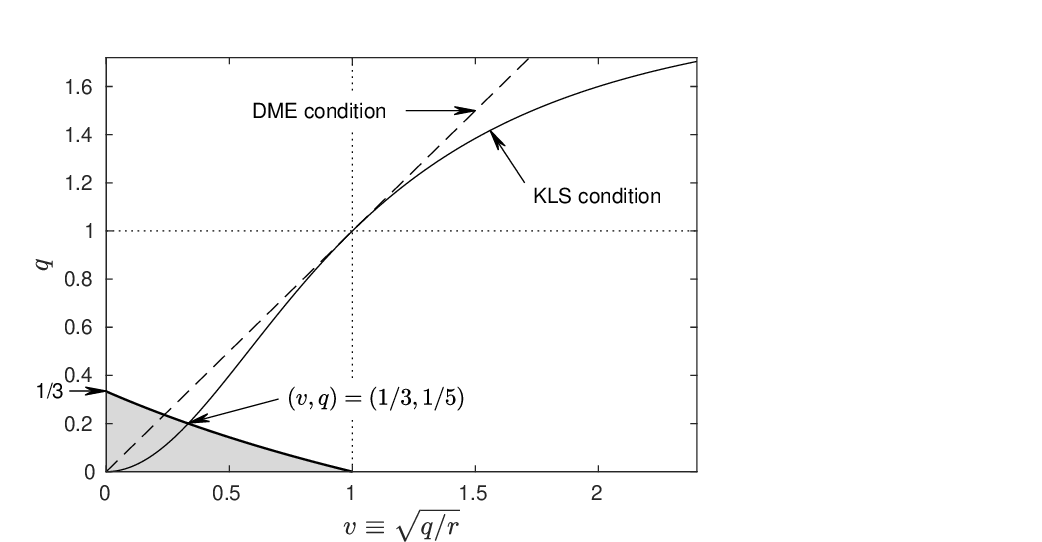}}
	\caption
	{
		Parameter plane (${q}$ vs ${v}$): 
		the shaded region is the one characterized by a bimodal current-density relation. 
		The thin dashed and solid lines respectively denote the DME and KLS conditions (see the text).
	}
	\label{fig:phase_diagram_v_q}
\end{figure}

In the framework of the PA theory, the current-density relation for the model considered here has already been obtained, in explicit analytical form, in various previous works \cite{MidhaKolomeiskyGupta2018jstat,PelizzolaPrettiPuccioni2019,PalGupta2021} (note that most of such papers actually deal with homogeneous rates, but this fact turns out to be irrelevant, due to the ``smoothness'' assumption).
Here however we report (in \ref{app:cont_lim}) a slightly more general derivation, which holds even in the absence of the symmetry assumption \eqref{eq:simmetria_parametri} (albeit we do not analyze this case further). 
As mentioned in the introduction, the physically relevant fact is that the current-vs-density diagram undergoes a transition from a unimodal shape, for attractive or weakly repulsive interactions, to a bimodal one, for stronger repulsive interactions.
Taking ${v}$ and ${q}$ as free parameters, the transition line between the two regimes, shown in figure~\ref{fig:phase_diagram_v_q}, is defined by the following equation 
\begin{equation}
	{q} = \frac{1-{v}}{3+{v}}
	\, . 
	\label{eq:phase_transition}
\end{equation}
Also the latter equation has already appeared in the literature\footnote{Equation (26) in \cite{PalGupta2021} precisely corresponds to our equation \eqref{eq:phase_transition}, whereas equation (17) in \cite{MidhaKolomeiskyGupta2018jstat} (seemingly quite different) actually includes some extra nonphysical solutions.}, but for the sake of completeness we report a derivation of it in \ref{app:ph_trans}. 
In the same figure~\ref{fig:phase_diagram_v_q} we report lines representing the DME and KLS conditions, respectively expressed by equations \eqref{eq:condizione_DME_vq} and \eqref{eq:condizione_KLS_vq}.
By comparing the latter equation with \eqref{eq:phase_transition} and eliminating ${q}$ we obtain the transition value for the interaction parameter, namely ${{v}=1/3}$.
As anticipated in the introduction, both aforementioned lines turn out to cross the transition line, and one can also observe that they are tangent to each other at point ${{q}={v}=1}$ (i.e.~for vanishing NN interaction). 
From now on we will always assume that equations \eqref{eq:condizioni_KLS} hold, and that accordingly the steady-state dynamics of the system are specified only by the interaction parameter ${v}$, obviously besides the rate modulation ratio $\lambda_{\max}/\lambda_{\min}$. 

\begin{figure}
	\flushright{\includegraphics*[width=0.85\textwidth]{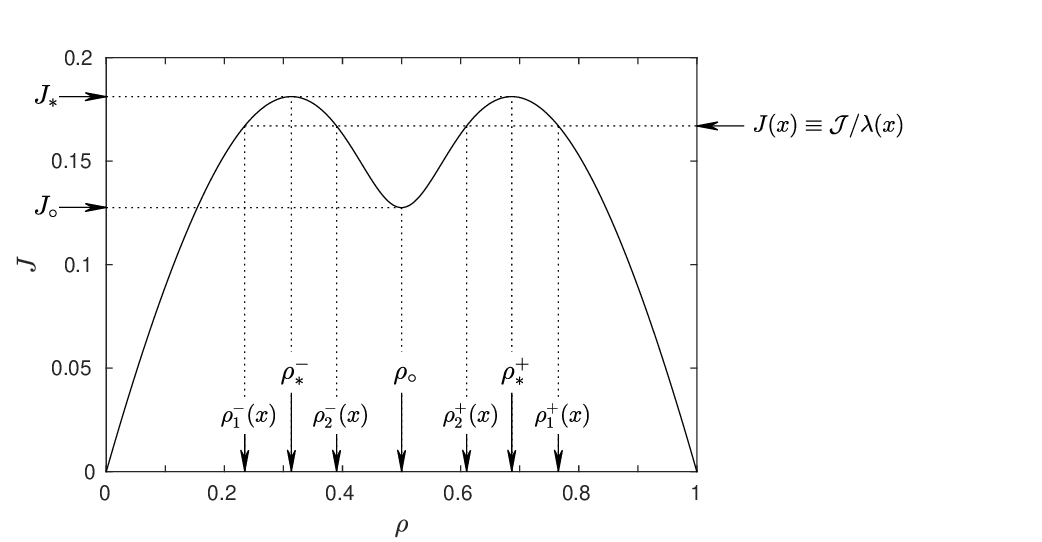}}
	\caption
	{
		Reduced current-density relation, evaluated with the symmetry assumption \eqref{eq:simmetria_parametri} and with the KLS condition \eqref{eq:condizioni_KLS}, for ${{v} = 0.15}$. 
	}
	\label{fig:diagramma_fondamentale}
\end{figure}

In figure \ref{fig:diagramma_fondamentale} we plot the (reduced) current-density relation ${{J} = {F}_{{p},{q},{r},{s}}(\rho)}$, evaluated in the hypotheses described above, namely \eqref{eq:simmetria_parametri} and \eqref{eq:condizioni_KLS}, for a given value of the interaction parameter ${v}$ in the bimodality region.
The explicit expression, parameterized by ${v}$ only: 
\begin{equation}
	{F}_{{v}}(\rho) = \frac{1}{1+{v}^2} \left[ 1 - 2 \frac{1-2\rho(1-\rho)}{1+\sqrt{1-4\rho(1-\rho)(1-{v}^2)}} \right]
	\, ,
\end{equation}
can be obtained by plugging \eqref{eq:cont_lim:definizione_I}, \eqref{eq:cont_lim:definizioni_rtilde_a_semplici} and \eqref{eq:cont_lim:correlaz7} into \eqref{eq:I_aafo_J:corrente1}, of course along with the KLS condition \eqref{eq:condizione_KLS_vq}. 
As previously mentioned, we can see that such relation is symmetric under the ${\rho \mapsto 1-\rho}$ transformation (mirror symmetry), as a consequence of \eqref{eq:simmetria_parametri}, and is characterized by a convex region around a ``central minimum'' at density ${\rho_\circ = 1/2}$ and two ``side maxima'' at densities ${\rho_*^\pm = 1/2 \pm \Delta}$. 
The function values corresponding to the (local) minimum and the (absolute) maxima, which we respectively denote by ${J}_\circ$ and ${J}_*$, are derived analytically in \ref{app:max_e_min}. 
Plugging the KLS condition \eqref{eq:condizione_KLS_vq} into \eqref{eq:max_e_min:Jzero} and \eqref{eq:max_e_min:Jstar}, we respectively get 
\begin{eqnarray}
	{J}_\circ & = \frac{ {v} }{ (1 + {v}^2) (1 + {v}) }
	\, , 
	\label{eq:Jzero_KLS}
	\\ 
	{J}_* & = \frac{ \left( \sqrt{2} - \sqrt{\vphantom{2} 1-{v}^2} \, \right)^2 }{ (1 + {v}^2) (1 + {v}) (1 - {v}) }
	\, . 
	\label{eq:Jstar_KLS}
\end{eqnarray}
We will see that the knowledge of these values allows one to determine, analytically, the transition between the two different regimes emerging in the strong interaction region. 
Figure \ref{fig:diagramma_fondamentale} also illustrates the meaning of equation \eqref{eq:profili_analitici}, which, given the reduced current ${J}(x)$, allows one to determine the possible density profiles, by inverting the current-density function (details in \ref{app:dens_prof}). 
It can be seen that, apart from degenerate cases, there can be either 2 or 4 possible density solutions, respectively for ${0<{J}(x)<{J}_\circ}$ or ${{J}_\circ<{ J}(x)<{J}_*}$. 
Due to the mirror symmetry, they turn out to be pairwise complementary, that is, according to the notation in figure~\ref{fig:diagramma_fondamentale}, 
\begin{subequations} 
\begin{eqnarray}
	\rho_{1}^{+}(x) = 1-\rho_{1}^{-}(x)
	\, , \\ 
	\rho_{2}^{+}(x) = 1-\rho_{2}^{-}(x)
	\, . 
\end{eqnarray}
\end{subequations}

\section{Density profiles}
\label{sec:density_profiles}

In this section we first describe the results that we have obtained at finite size, by the numerical PA method, introduced in the previous one. 
As mentioned above, at the size considered (${{L}=10000}$) the smooth sections of the density profiles turn out to be practically indistinguishable from the analytical solutions, obtained in the continuum limit. 
We consider two different combinations for the relevant model parameters, representative of the two different regimes that we have observed. 
In particular, keeping the rate modulation ratio fixed at ${\lambda_{\max}/\lambda_{\min} = 1.5}$, we take two different values of the interaction parameter, namely ${{v}=0.15}$ and ${{v}=0.10}$ (recall that smaller ${v}$ means stronger repulsion). 
We respectively denote the two regimes as multi-shock (MS) and small-shock (SS), for reasons that will be immediately clear from the description. 

\begin{figure}
	\flushright{\includegraphics*[width=0.85\textwidth]{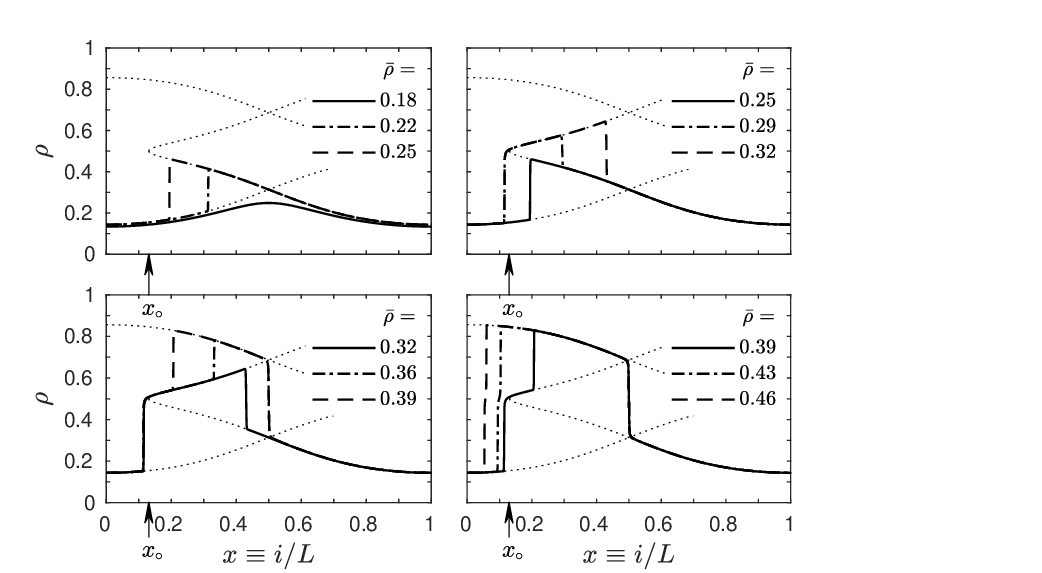}}
	\caption{
		MS regime: ${\lambda_{\max}/\lambda_{\min} = 1.5}$, ${{v} = 0.15}$. 
		Different line types denote density profiles, computed by the PA method at finite but large size (${{L} = 10000}$), for different values of the mean density $\bar{\rho}$ (see the legends). 
		Thin dotted lines denote the analytical solutions \eqref{eq:profili_analitici} at maximal current ${\Jcal_{\max} = \lambda_{\min} {J}_*}$. 
	}
	\label{fig:profili_multishock}
\end{figure}

The MS regime is illustrated in figure~\ref{fig:profili_multishock}. 
For low enough values of the mean density $\bar{\rho}$, the system exhibits a smooth profile, with a density maximum at position ${x_{\min}=1/2}$, corresponding to the minimum of the rate modulation function \eqref{eq:rmf}. 
We denote this phase as the LD (low-density) phase. 
Let us now imagine to increase $\bar{\rho}$. 
At a certain value (roughly ${\bar{\rho} \approx 0.20}$), the steady-state profile begins to develop a shock, initially placed at $x_{\min}$, with vanishing amplitude, and progressively moving backward (i.e.~towards smaller $x$ values, opposite to the particle flux), with increasing amplitude. 
We denote the latter phase, characterized by 1 shock, as S1L (where L stands for \emph{low} density). 
Such a displacement of the shock position for increasing $\bar{\rho}$ goes on until it reaches a peculiar position, say $x_\circ$. 
After that, a further increase of $\bar{\rho}$ no longer affects this first shock, but it induces the onset of a second shock, initially placed at $x_\circ$ (with vanishing amplitude) and progressively moving forward (with increasing amplitude). 
The transition to this new phase, characterized by 2 shocks (and denoted as S2L), occurs roughly at ${\bar{\rho} \approx 0.27}$.  
Once again, the displacement of the second shock goes on until it reaches another peculiar position, in this case coinciding with $x_{\min}$. 
Upon further increasing $\bar{\rho}$, even this shock remains ``locked'', while a third shock sets on, initially placed at $x_{\min}$ (with vanishing amplitude) and subsequently moving backward (with increasing amplitude). 
The transition to the latter 3-shock phase (denoted as S3L) occurs roughly at ${\bar{\rho} \approx 0.34}$.  
Furthermore, around ${\bar{\rho} \approx 0.41}$, the third shock joins the first one at $x_\circ$, giving rise to a single shock (with double amplitude), which still 
moves backward for increasing $\bar{\rho}$. 
Note that the shock placed at $x_{\min}$ is still there, so that we have another 2-shock phase (denoted as S2), which extends over a range of average densities including the half-filling value ${\bar{\rho} = 0.5}$. 
Needless to remark that all observed shocks are properties of the steady state, so what we call ``displacements'' do not happen over time but are actually changes in the steady state of the system, in response to changes in its mean density $\bar{\rho}$. 
For ${\bar{\rho} > 0.5}$ the behaviour of the system can be argued from the particle-hole symmetry. 
In particular we can see that transforming the average density as ${\bar{\rho} \mapsto 1-\bar{\rho}}$ entails the following transformation for the density profiles 
\begin{equation}
	\rho(x) \ \longmapsto \ 1-\rho(1-x)
	\, . 
\end{equation}
As a consequence, it is possible to identify the high-density counterparts of the various shock phases S1L, S2L, S3L, which we respectively denote as S1H, S2H, S3H.  
Of course, also a smooth high-density phase exists, which we denote as HD. 
As far as the current is concerned, it can be observed that it keeps a constant maximum value in all the shock phases, and that the value computed at finite size is practically indistinguishable from the continuum-limit value 
\begin{equation}
	\Jcal_{\max} = \lambda_{\min} {J}_{*}
	\, , 
	\label{eq:maximal_current}
\end{equation}
where we recall that ${J}_{*}$ denotes the maximum of the reduced current-density relation (figure~\ref{fig:diagramma_fondamentale}). 
Looking at figure~\ref{fig:diagramma_fondamentale}, one can also argue that the $x_\circ$ position, introduced above, is the one where the intermediate-density solutions become degenerate, namely ${\rho_2^\pm(x_\circ) = \rho_\circ = 1/2}$, and the reduced current takes value ${{J}(x_\circ) = {J}_\circ}$ (with a slight abuse of language, we will call this the \emph{critical point}). 
As a consequence, from equations \eqref{eq:reduced_current} and \eqref{eq:maximal_current}, with ${\Jcal = \Jcal_{\max}}$, we can write the equation for $x_\circ$ as 
\begin{equation}
	\frac{\lambda(x_\circ)}{\lambda_{\min}} = \frac{{J}_*}{{J}_\circ} 
	\, ,
	\label{eq:equazione_xzero}
\end{equation} 
where we recall that ${J}_\circ$ and ${J}_*$ are known analytically from \eqref{eq:Jzero_KLS} and \eqref{eq:Jstar_KLS} (see also \ref{app:max_e_min}). 
Taking into account the specific form \eqref{eq:rmf} of the rate modulation function $\lambda(x)$, we obtain  
\begin{equation}
	x_\circ = \frac{1}{2 \pi} \arccos \left( 1 - 2 \, \frac{\lambda_{\max}/\lambda_{\min} - {J}_*/{J}_\circ}{\lambda_{\max}/\lambda_{\min} - 1} \right) 
	\, .
\end{equation} 
Let us note, in figure~\ref{fig:profili_multishock}, that the shock we claim to be placed at $x_\circ$, in the S2L and S3L phases, is in fact slightly displaced to the left, and the same goes for the S3L/S2 transition, where the two-shock merging appears to occur at $x$ slightly lower than $x_\circ$. 
Even though we do not display the results, we have a quite clear evidence that both such discrepancies are finite-size effects, since they tend to vanish upon increasing the number of nodes ${L}$. 
In fact, it is not surprising that finite size effects are especially relevant in the neighbourhood of the critical point, where the derivative of $\rho(x)$ diverges. 

\begin{figure}
	\flushright{\includegraphics*[width=0.85\textwidth]{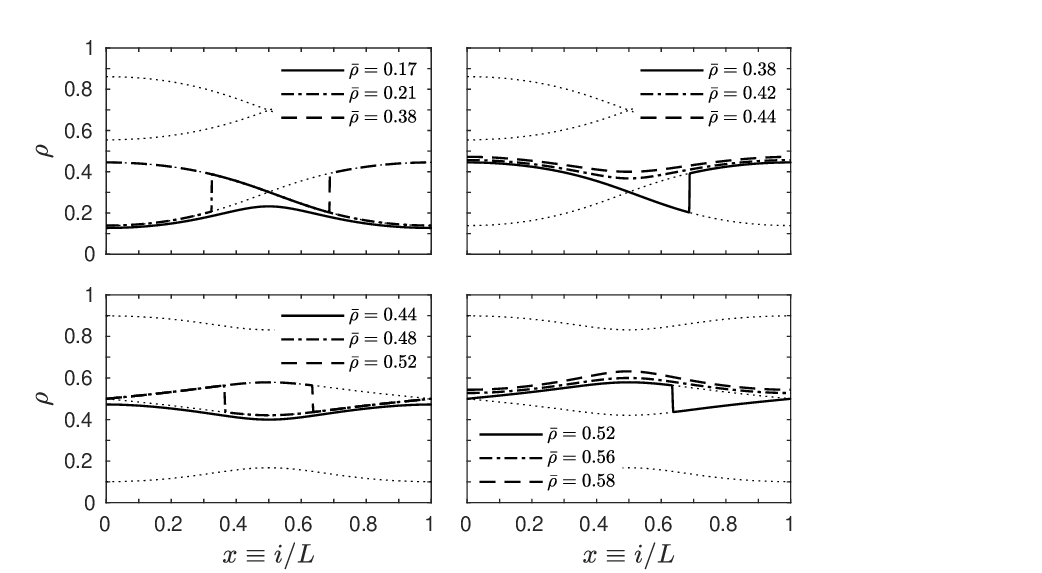}}
	\caption
	{
		SS regime: ${\lambda_{\max}/\lambda_{\min} = 1.5}$, ${{v} = 0.10}$.
		Thin dotted lines denote the analytical solutions \eqref{eq:profili_analitici} at maximal current ${\Jcal_{\max} = \lambda_{\min} {J}_*}$ (top panels) or minimal current ${\Jcal_{\min} = \lambda_{\max} {J}_\circ}$ (bottom panels). 
		Other lines as in figure~\ref{fig:profili_multishock}.
	}
	\label{fig:profili_smallshock}
\end{figure}

Let us now observe that, if the rate modulation ratio is not large enough, specifically 
\begin{equation}
	\frac{\lambda_{\max}}{\lambda_{\min}} < \frac{{J}_*}{{J}_\circ} 
	\, ,
	\label{eq:regime_SS}
\end{equation} 
then equation \eqref{eq:equazione_xzero} has no more solutions, meaning that no critical point can exist. 
Keeping the left-hand side fixed, the above condition can also occur if the right-hand side becomes exceedingly large, as a consequence of a stronger repulsive interaction. 
In both cases, a very different transition scenario takes place (SS regime), which we illustrate in figure~\ref{fig:profili_smallshock}. 
For low values of the mean density $\bar{\rho}$, the system still exhibits a smooth profile (LD phase), which undergoes a transition to the 1-shock phase (S1L) upon increasing $\bar{\rho}$ (the transition occurs roughly at ${\bar{\rho} \approx 0.19}$). 
As in the MS regime, the shock is initially placed at ${x_{\min}=1/2}$, with vanishing amplitude, and, upon further increasing $\bar{\rho}$, it progressively moves backward with increasing amplitude. 
Here begins the difference with respect to the MS regime. 
Due to the lack of a critical point, the shock is allowed to go through the entire system, so that in figure~\ref{fig:profili_smallshock} we see it exiting our ``observation window'' at ${x=0}$ and reentering at ${x=1}$. 
At this point, corresponding to the maximum of the rate modulation function \eqref{eq:rmf}, the shock takes maximum amplitude, and subsequently proceeds backward, with decreasing amplitude. 
For ${\bar{\rho} \approx 0.40}$, it finally comes back to $x_{\min}$ and disappears, giving rise to another phase featuring a smooth profile, which we denote as LD'. 
In the whole S1L phase the current takes the constant maximal value \eqref{eq:maximal_current} (apart from the usual vanishing discrepancy, due to finite size), whereas in the LD' phase the current turns out to decrease upon increasing $\bar{\rho}$. 
Note that the latter is a typical feature of a high-density phase, even though in this case it takes place in a low-density range, i.e.~${\bar{\rho} < 0.5}$.     
This is clearly related to the presence of side maxima (specifically the one at lower density) in the current-density relation. 
Upon further increasing the mean density, roughly above ${\bar{\rho} \approx 0.46}$, we can observe the onset of another shock, which in this case originates at ${x = 0}$, that is, at the \emph{maximum} of the rate modulation function. 
The shock moves \emph{forward} on increasing $\bar{\rho}$, it goes trough the whole system, with maximum amplitude at $x_{\min}$, and finally disappears at its initial position (${x = 1}$ in our view). 
We denote this last 1-shock phase as S1', in order to distinguish it from the ``normal'' 1-shock phase (S1), occurring when the current-density relation is unimodal (that is, in case of weak \cite{PalGupta2021} or lacking \cite{BanerjeeBasu2020} NN interaction). 
A first peculiar feature of the S1' phase is the \emph{descending} shock (i.e. going from higher to lower density), which does not appear in any other 1-shock phase (including S1L and S1H). 
Another one is that the current is constant (i.e.~independent of $\bar{\rho}$) but \emph{minimal}, specifically taking value 
\begin{equation}
	\Jcal_{\min} = \lambda_{\max} {J}_{\circ}
	\, , 
	\label{eq:minimal_current}
\end{equation}
where we recall that ${J}_{\circ}$ denotes the minimum of the current-density relation (figure~\ref{fig:diagramma_fondamentale}). 
In fact, the onset of a minimal-current phase is related to the presence of the central minimum in the current-density relation, as recognized for instance in \cite{PopkovSchutz1999,HagerKrugPopkovSchutz2001}, and it has actually been observed in several similar models. 
For ${\bar{\rho} > 0.5}$ the behaviour of the system is still constrained by the particle-hole symmetry, so that, as done in the MS regime, it is possible to identify the high-density counterparts of the phases encountered at ${\bar{\rho} < 0.5}$, which we respectively denote as HD, S1H, HD' (the S1' phase extends roughly up to ${\bar{\rho} \approx 0.54}$). 
The symmetry also entails that in the HD' phase the current increases upon increasing $\bar{\rho}$, which is actually a typical feature of a low-density phase. 

We conclude this section by discussing the fact that, as already mentioned in section~\ref{sec:methods}, in principle it is possible to completely determine the density profiles in the continuum limit, even without making use of the numerical PA method, but using only the analytical solutions of equation~\eqref{eq:profili_analitici}.
Figures \ref{fig:profili_multishock} and \ref{fig:profili_smallshock} clearly confirm that, in the shock phases, the smooth sections of the density profiles always match with the analytical profiles evaluated at maximal or minimal current.
In all the observed phases, including the MS regime, there is always only one shock whose position, say $x_\mathrm{s}$, depends on the mean density of the system, while any other shock is ``locked''. 
To determine $x_\mathrm{s}$, we need to solve equation~\eqref{eq:densita_media} (the integral must be done numerically), for a suitable profile $\rho(x)$, being the union of known continuous-profile sections, with a discontinuity at $x_\mathrm{s}$.
We refer to section~\ref{sec:numerical_simulations} for more details about the specific equations for the various phases.
Here we just discuss the fact that such equations generally have two possible solutions, and how the unphysical one can be detected.
In our case, the matter is specially simple, since the smooth profiles are characterized by a mirror symmetry with respect to ${x_{\min}=1/2}$ (following from the same property of the rate modulation function), which is broken by shock profiles. 
As a consequence, if $\rho(x)$ is a solution of equation~\eqref{eq:densita_media}, the ``mirrored'' profile $\rho(1-x)$ is also a solution, and therefore a shock placed in $x_\mathrm {s}$ in the former profile, is shifted to ${1-x_\mathrm{s}}$ in the latter. 
Furthermore, if the shock is ``ascending'' in the former profile, it becomes ``descending'' in the latter (and vice versa), and consequently one is stable and the other unstable, so that ultimately only one of the two profiles can be the stationary one. 
Let us recall that the stability of a shock depends on the propagation velocity of small perturbations of the density profile, i.e.~on the derivative of the current-density relation (the so-called collective velocity) \cite{PopkovSchutz1999,HagerKrugPopkovSchutz2001,KolomeiskySchutzKolomeiskyStraley1998}. 
In particular, stability occurs if the propagations on the two sides of the shock are opposite and directed towards the shock itself (see in particular equation~(9) in~\cite{HagerKrugPopkovSchutz2001}). 
By comparing figures \ref{fig:profili_multishock} and \ref{fig:profili_smallshock} with the current-density relation in figure~\ref{fig:diagramma_fondamentale}, it can be verified that the profiles obtained by the numerical PA method always satisfy the stability criterion. 
The aforementioned ``locked shocks'', observed in 2- and 3-shock phases, are partly an exception, and can be considered marginally stable, as they involve at least one point of the current-density function with zero derivative. 
In any case, it can be seen that the numerical PA method is convenient also in this respect, since, by following the dynamical evolution of the system, it ``automatically'' determines the stable stationary state.

\section{Phase diagrams} 
\label{sec:phase_diagrams}

In this section we summarize some results which have already been outlined in the previous one, but which can be described more precisely and exhaustively in terms of phase diagrams.

\begin{figure}
	\flushright{\includegraphics*[width=0.85\textwidth]{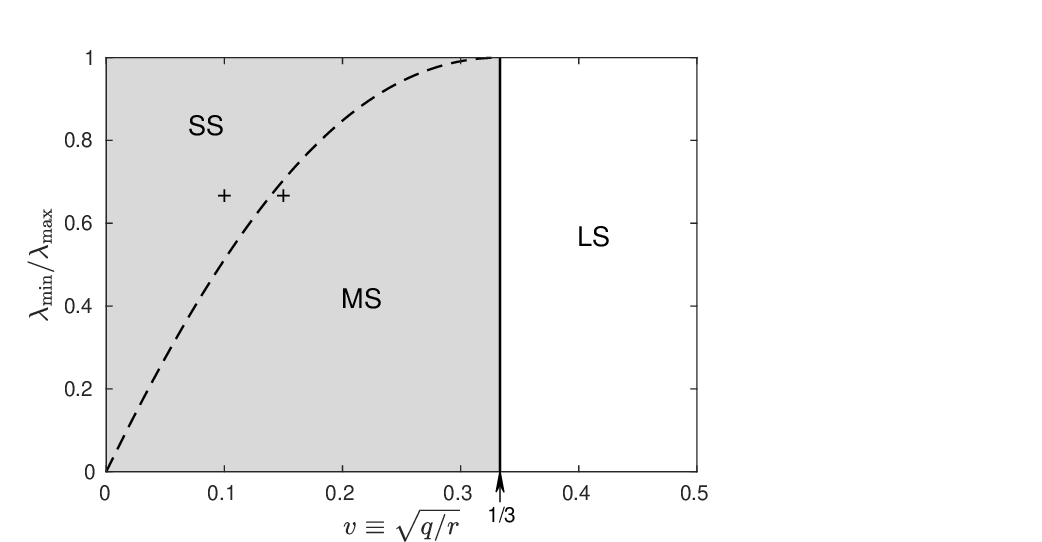}}
	\caption
	{
		Reciprocal rate-modulation ratio ($\lambda_{\min}/\lambda_{\max}$) vs interaction parameter (${v}$) phase diagram. 
		Tags denote different regimes, described in the text. 
		The transition line between the MS and SS regimes (dashed line) is defined by equation~\eqref{eq:transizione_MS-SS}. 
		The shaded region is the one characterized by a bimodal current-density relation. 
		Cross symbols mark the particular cases analyzed in the previous section (${\lambda_{\min}/\lambda_{\max}=2/3}$, ${{v}=0.10,0.15}$). 
	}
	\label{fig:phase_diagram_v_z}
\end{figure}

First, we have seen that the interesting behaviours (i.e.~the two regimes tagged MS and SS) occur when the current-density relation is bimodal, for a sufficiently strong repulsive interaction (i.e.~for small enough values of the interaction parameter: ${{v}<1/3}$).
For ${{v}>1/3}$ the phenomenology is qualitatively equivalent to that of a system without NN interaction, where a single shock phase (S1) occurs~\cite{BanerjeeBasu2020}.
We call this the large-shock (LS) regime, since the (unique) shock, occurring between a low-density (LD) and a high-density (HD) phase, can assume greater amplitudes than those observed in the MS and SS regimes.
The transition between the latter two regimes is also controlled by another parameter, namely the rate-modulation ratio $\lambda_{\max}/\lambda_{\min}$.
The transition line is determined by~\eqref{eq:regime_SS} taken as an equality, where the right-hand side depends on ${v}$ according to \eqref{eq:Jzero_KLS} and \eqref{eq:Jstar_KLS} (see also \ref{app:max_e_min}), thence  
\begin{equation}
	\frac{\lambda_{\max}}{\lambda_{\min}}
	= \frac{ \left( \sqrt{2} - \sqrt{\vphantom{2} 1-{v}^2} \, \right)^2 }{ {v} \, (1 - {v}) }
	\, .
	\label{eq:transizione_MS-SS}
\end{equation}
The resulting phase diagram is shown in figure~\ref{fig:phase_diagram_v_z}, where for clarity we have also marked the specific points analyzed in the previous section. 

\begin{figure}
	\flushright{\includegraphics*[width=0.85\textwidth]{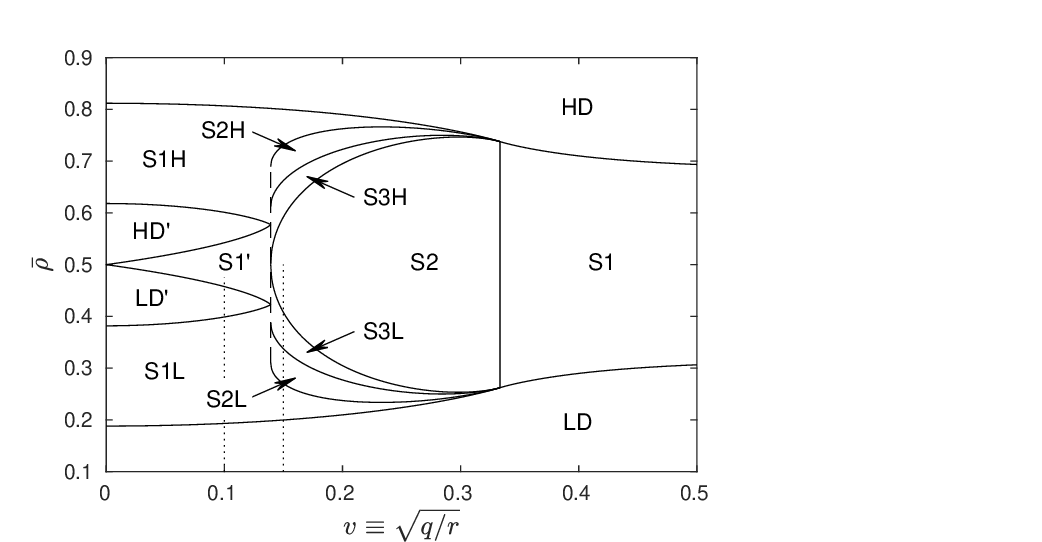}}
	\caption
	{
		Phase diagram at constant rate-modulation ratio (${\lambda_{\max}/\lambda_{\min} = 1.5}$): mean density $\bar{\rho}$ vs interaction parameter ${v}$.
		Solid lines denote continuous transitions. 
		A dashed line denotes the crossover regime between MS and SS (see the text). 
		Phase tags are also explained in the text.
		Thin dotted lines mark the particular cases analyzed in the previous section (${{v}=0.10,0.15}$). 
	}
	\label{fig:phase_diagram_v_rhomedia}
\end{figure}

\begin{figure}
	\flushright{\includegraphics*[width=0.85\textwidth]{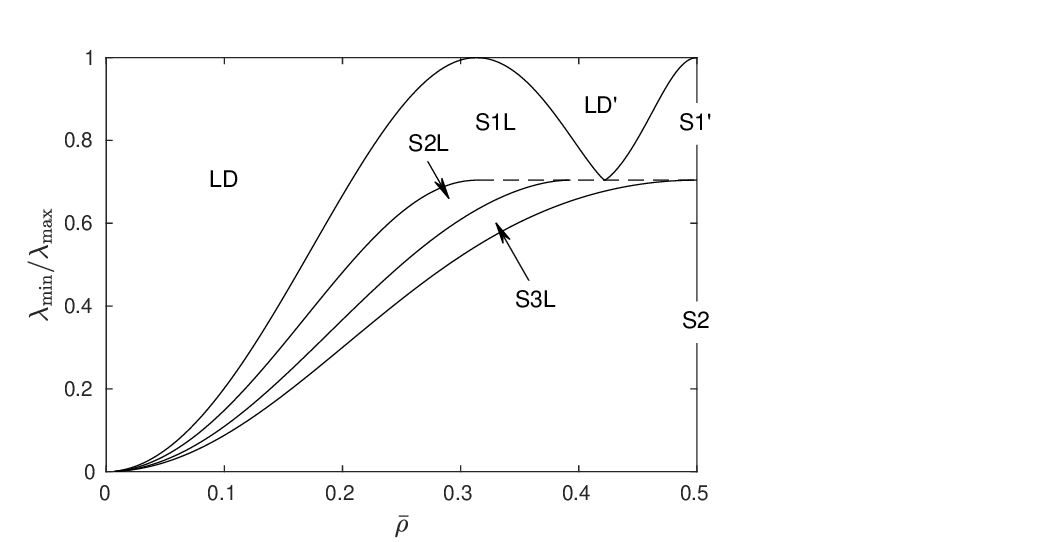}}
	\caption
	{
		Phase diagram at constant interaction parameter (${{v} = 0.15}$): reciprocal rate-modulation ratio $\lambda_{\min}/\lambda_{\max}$ vs mean density $\bar{\rho}$.
        Lines and tags as in figure~\ref{fig:phase_diagram_v_rhomedia}. 
	}
	\label{fig:phase_diagram_rhomedia_z}
\end{figure}

By varying the mean density $\bar{\rho}$, in the previous section we pointed out two different sequences of phase transitions (characterizing the MS and SS regimes) and we approximately identified the $\bar{\rho}$ values at which the transitions take place.
Taking into account the analytical profiles obtained from the continuum PA theory (section~\ref{sec:methods}), it is possible to precisely determine these threshold values, and also their evolution, in response to arbitrary variations of the control parameters.
Let us first consider the MS regime and in particular the density profiles occurring at the 4 transitions, observed in the previous section. 
We can see that, limited to the ${(x_{\circ},x_{\min})}$ interval, such profiles correspond respectively to the 4 possible analytical solutions, which we have denoted as $\rho_1^-( x)$, $\rho_2^-(x)$, $\rho_2^+(x)$, $\rho_1^+(x)$ (in increasing order of density). 
Moreover, in the remaining intervals ${[0,x_\circ)}$ and ${(x_{\min},1)}$ (which, taking into account the periodicity of the system, are effectively equivalent to a single interval ${(x_{\min}-1,x_\circ)}$), all the transition profiles match with the lowest-density solution $\rho_1^-(x)$.
Thus, generically denoting with $\bar{\rho}_\mathrm{A/B}$ the mean density at which the transition between phases $\mathrm{A}$ and $\mathrm{B}$ takes place, we can write\footnote{In the first equation, splitting into two integrals is unnecessary, but it better highlights the analogy with subsequent ones.} 
\begin{subequations} 
\label{eq:transizioni_MS} 
\begin{eqnarray}
	\bar{\rho}_{\mathrm{LD/S1L}}  & = \int_{x_{\min}-1}^{x_\circ} \left. {\rho}_{1}^{-}(x) \right|_{\Jcal_{\max}} \rmd x  +  \int_{x_\circ}^{x_{\min}} \left. {\rho}_{1}^{-}(x) \right|_{\Jcal_{\max}} \rmd x
	\label{eq:transizione_MS} 
	\, , \\
	\bar{\rho}_{\mathrm{S1L/S2L}} & = \int_{x_{\min}-1}^{x_\circ} \left. {\rho}_{1}^{-}(x) \right|_{\Jcal_{\max}} \rmd x  +  \int_{x_\circ}^{x_{\min}} \left. {\rho}_{2}^{-}(x) \right|_{\Jcal_{\max}} \rmd x
	\, , \\
	\bar{\rho}_{\mathrm{S2L/S3L}} & = \int_{x_{\min}-1}^{x_\circ} \left. {\rho}_{1}^{-}(x) \right|_{\Jcal_{\max}} \rmd x  +  \int_{x_\circ}^{x_{\min}} \left. {\rho}_{2}^{+}(x) \right|_{\Jcal_{\max}} \rmd x
	\, , \\
	\bar{\rho}_{\mathrm{S3L/S2}}  & = \int_{x_{\min}-1}^{x_\circ} \left. {\rho}_{1}^{-}(x) \right|_{\Jcal_{\max}} \rmd x  +  \int_{x_\circ}^{x_{\min}} \left. {\rho}_{1}^{+}(x) \right|_{\Jcal_{\max}} \rmd x
	\, ,
\end{eqnarray} 
\end{subequations} 
where an appropriate subscript reminds us that the profiles are always calculated at maximum current. 
As far as the SS regime is concerned, it can be seen that the profiles occuring at the 3 observed transitions correspond entirely to single analytical solutions, which however can be at either maximum or minimum current. 
In particular, with the same notation used above, we have 
\begin{subequations} 
\label{eq:transizioni_SS} 
\begin{eqnarray}
	\bar{\rho}_{\mathrm{LD/S1L}}  & = \int_{0}^{1} \left. {\rho}_{1}^{-}(x) \right|_{\Jcal_{\max}} \rmd x
	\label{eq:transizione_SS} 
	\, , \\
	\bar{\rho}_{\mathrm{S1L/LD'}} & = \int_{0}^{1} \left. {\rho}_{2}^{-}(x) \right|_{\Jcal_{\max}} \rmd x
	\, , \\
	\bar{\rho}_{\mathrm{LD'/S1'}} & = \int_{0}^{1} \left. {\rho}_{2}^{-}(x) \right|_{\Jcal_{\min}} \rmd x
	\, ,
\end{eqnarray} 
\end{subequations} 
where we note that, still due to periodicity, \eqref{eq:transizione_SS} is equivalent to \eqref{eq:transizione_MS}. 
All the above integrals can be solved numerically with negligible computational effort, just because the integrand functions are all known analytically.
Keeping the rate-modulation ratio fixed and varying the interaction parameter, we obtain the phase diagram shown in figure~\ref{fig:phase_diagram_v_rhomedia}, where again we have marked (thin dotted lines) the two cases studied in the previous section. 
Note that the diagram has been completed by symmetry, adding the complementary phases that can be observed in the high density range (${\bar{\rho}>1/2}$). 
On the other hand, keeping the interaction parameter fixed and varying the rate-modulation ratio, we obtain the phase diagram shown in figure~\ref{fig:phase_diagram_rhomedia_z}. 
Let us finally observe that, in both figures \ref{fig:phase_diagram_v_rhomedia} and \ref{fig:phase_diagram_rhomedia_z}, almost all transitions (namely, all those denoted by solid lines) are continuous. 
In fact they are characterized by the appearance or disappearance of a shock, whose amplitude does not exhibit any discontinuity, as a function of the control parameters. 
The only exception is the transition denoted by the dashed line, which occurs simultaneously with the crossover between the MS and SS regimes. 
We refer to the next section for a closer discussion of this ``crossover regime''. 
By now we only mention the fact that, in this peculiar situation, the critical point $x_\circ$ becomes degenerate with its complementary ${1-x_\circ}$ (we shall denote such phenomenon as \emph{coalescence}), which entails that the shock occurring at that point in the S2L and S3L phases (and similarly in S2H and S3H) is no longer locked. 
As a result, the mean-density constraint equation \eqref{eq:densita_media} is no longer sufficient to determine the locations of all shocks. 
It is reasonable to expect that physically this corresponds to the onset of so-called delocalized and synchronized domain walls~\cite{BanerjeeBasu2020}, i.e.~two shocks whose positions are indeed stochastic processes (random walks), yet ``rigidly'' bound to each other because of equation~\eqref{eq:densita_media}.

\section{Numerical simulations}
\label{sec:numerical_simulations}

The main purpose of this section is to provide quite robust evidence that, under the KLS condition, the PA theory in the continuum limit describes the behavior of the system in a very accurate (plausibly exact) manner, thus also confirming the accuracy (or even the exactness) of the phase diagrams shown in the previous section.
To do this we are going to compare the density profiles obtained from kinetic Monte Carlo (KMC) simulations (with increasing sizes: ${{L}=2000,5000,10000}$) with those obtained from the analytical theory (in the ${{L}\to\infty}$ limit).
We focus on the shock phases, since in the smooth phases an even better matching is obtained, at relatively small sizes, of the order of ${{L}=500,1000}$. 
The simulations are carried out using the well-established Gillespie algorithm. 
As usual, each simulation is first run for a ``settling time'' $t_{\mathrm{set}}$, to ensure that the system relaxes to steady state, after which averages are computed over an ``averaging time'' $t_{\mathrm{ave}}$. 
These characteristic times have been empirically adjusted, leading us to choose (at the largest system size considered)
${t_{\mathrm{set}} = 2 \times 10^5}$ and ${t_{\mathrm{ave}} = 10^6}$ (the time unit is fixed by ${\lambda_{\min} = 1}$). 

As far as the analytical profiles are concerned, we recall that just one shock position varies with the mean density, and that such position must be determined according to the scheme outlined at the end of section~\ref{sec:density_profiles}, i.e.~taking into account both the density equation~\eqref{eq:densita_media} and the stability criterion. 
As previously mentioned, the results of the numerical PA method help us to select one of two possible profile shapes, on which the only constraint equation~\eqref{eq:densita_media} is left to be imposed.
The equations can be conveniently written, making use of the mean-density transition values \eqref{eq:transizioni_MS} and \eqref{eq:transizioni_SS}, and evaluating the corrections that occur at a given mean density $\bar{\rho}$ as a function of the shock position. 
In particular, for the different phases observed in the MS regime we have
\begin{subequations} 
\begin{eqnarray}
	\int_{x_{\mathrm{S1L}}}^{x_{\min}} \left[ {\rho}_{2}^{-}(x) - {\rho}_{1}^{-}(x) \right]_{\Jcal_{\max}} \rmd x 
	& = \bar{\rho} - \bar{\rho}_{\mathrm{LD/S1L}} 
	\label{eq:shock_position}
	\, , \\
	\int_{x_{\mathrm{S2L}}}^{x_{\min}} \left[ {\rho}_{2}^{-}(x) - {\rho}_{2}^{+}(x) \right]_{\Jcal_{\max}} \rmd x 
	& = \bar{\rho} - \bar{\rho}_{\mathrm{S2L/S3L}} 
	\, , \\
	\int_{x_{\mathrm{S3L}}}^{x_{\min}} \left[ {\rho}_{1}^{+}(x) - {\rho}_{2}^{+}(x) \right]_{\Jcal_{\max}} \rmd x 
	& = \bar{\rho} - \bar{\rho}_{\mathrm{S2L/S3L}} 
	\, , \\
	\int_{x_{\mathrm{S2 }}}^{x_{\min}} \left[ {\rho}_{1}^{+}(x) - {\rho}_{1}^{-}(x) \right]_{\Jcal_{\max}} \rmd x 
	& = \bar{\rho} - \bar{\rho}_{\mathrm{LD/S1L}} 
	\, ,
\end{eqnarray} 
\end{subequations} 
where we see that the generic unknown $x_\mathrm{A}$, denoting the shock position in a given phase $\mathrm{A}$, appears solely as lower bound of the integration interval. 
Regarding the SS regime, in the S1L phase the equation is equivalent to \eqref{eq:shock_position}, whereas in the S1' phase we have 
\begin{equation}
	\int_{0}^{x_{\mathrm{S1'}}} \left[ {\rho}_{2}^{+}(x) - {\rho}_{2}^{-}(x) \right]_{\Jcal_{\min}} \rmd x 
	= \bar{\rho} - \bar{\rho}_{\mathrm{LD'/S1'}} 
	\, .
\end{equation}

\begin{figure}
	\flushright{\includegraphics*[width=0.85\textwidth]{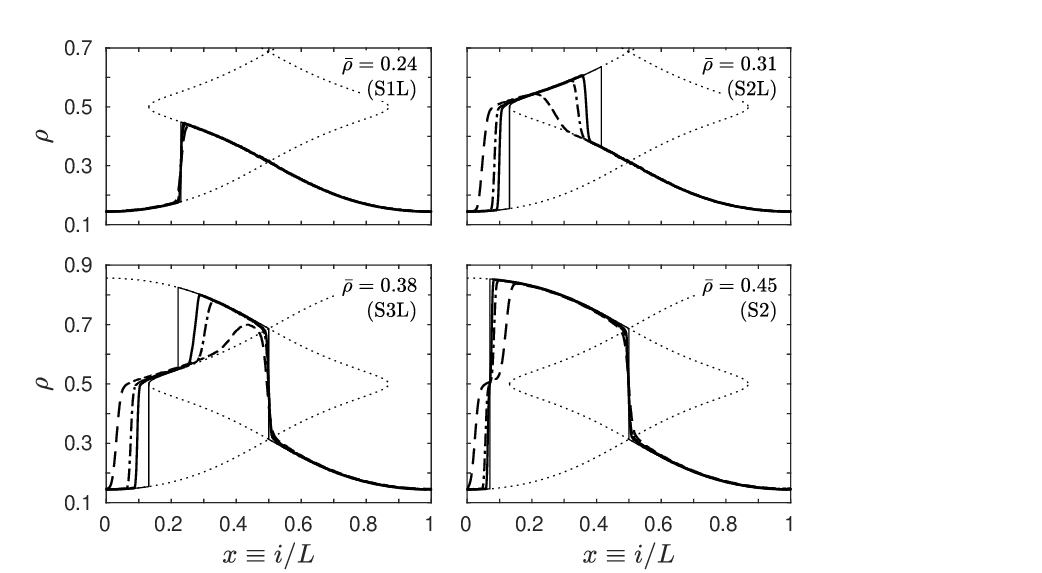}}
	\caption
	{
		MS regime (${\lambda_{\max}/\lambda_{\min} = 1.5}$, ${{v} = 0.15}$), for different values of the mean density: ${\bar{\rho}=0.24}$ (S1L phase), $0.31$ (S2L phase), $0.38$ (S3L phase), $0.45$ (S2 phase). 
		Thin solid lines denote density profiles, obtained by the continuum PA theory. 
		Dotted lines denote all possible solutions of \eqref{eq:profili_analitici} at maximal current ${\Jcal_{\max} = \lambda_{\min} {J}_*}$. 
		Thicker lines denote corresponding KMC simulation results for ${L=2000}$ (dashed lines), $5000$ (dash-dotted lines), $10000$ (solid lines). 
	}
	\label{fig:profili_multishock_simulati}
\end{figure}
\begin{figure}
	\flushright{\includegraphics*[width=0.85\textwidth]{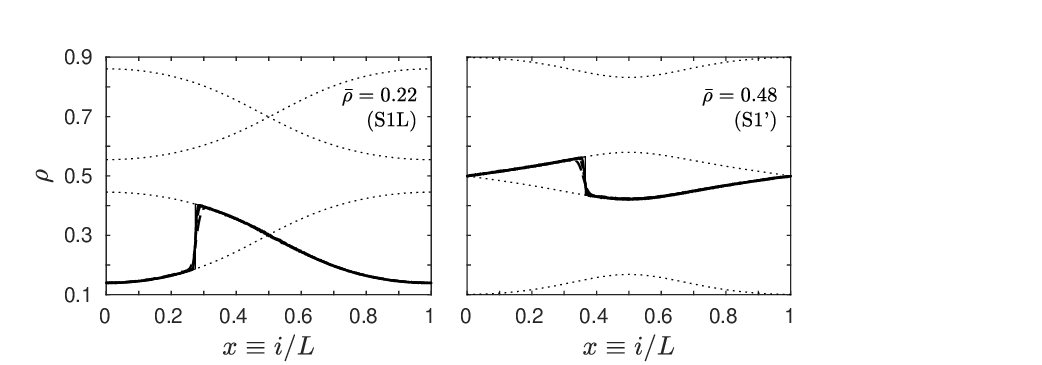}}
	\caption
	{
		SS regime: (${\lambda_{\max}/\lambda_{\min} = 1.5}$, ${{v} = 0.10}$), for different values of the mean density: ${\bar{\rho}=0.22}$ (S1L phase), $0.48$ (S1' phase). 
		Dotted lines denote all possible solutions of \eqref{eq:profili_analitici} at maximal current ${\Jcal_{\max} = \lambda_{\min} {J}_*}$ (left panel) or minimal current ${\Jcal_{\min} = \lambda_{\max} {J}_\circ}$ (right panel). 
		Other lines as in figure~\ref{fig:profili_multishock_simulati}.
	}
	\label{fig:profili_smallshock_simulati}
\end{figure}
The results, relating to the MS and SS regimes, are shown respectively in figures \ref{fig:profili_multishock_simulati} and~\ref{fig:profili_smallshock_simulati}. 
First of all, one observes that, as previously mentioned, the smooth sections of the simulated profiles match perfectly with the analytical solutions.
Also, in the shock phases, two considerably different situations are observed, depending on whether only one shock or more than one are there.
In the former case, i.e.~in phases S1L and S1' (figures \ref{fig:profili_multishock_simulati} and~\ref{fig:profili_smallshock_simulati}), there is a clear convergence of the simulation results towards the analytical ones, as the system size increases. 
Conversely, in the latter case, i.e.~in phases S2L, S3L and S2 (figure~\ref{fig:profili_multishock_simulati}), although the expected trend is still quite evident, extremely relevant finite-size effects occur, such that, even at the maximum size considered, some shocks are still quite far from the theoretical prediction. 
As already noticed in section~\ref{sec:density_profiles} about the numerical PA results, the critical point seems to play a key role in these effects.
Indeed, especially in phases S2L and S3L, the largest discrepancy between theory and simulation is observed for the shock lying in the vicinity of this point, whereas the similar discrepancy for the nearby (rightward) shock can be interpreted just as a side effect of the former, mediated by the mean-density constraint~\eqref{eq:densita_media}. 
Needless to recall that, due to symmetry, all the above remarks could be analogously repeated for the respective high-density phases.

\begin{figure}
	\flushright{\includegraphics*[width=0.85\textwidth]{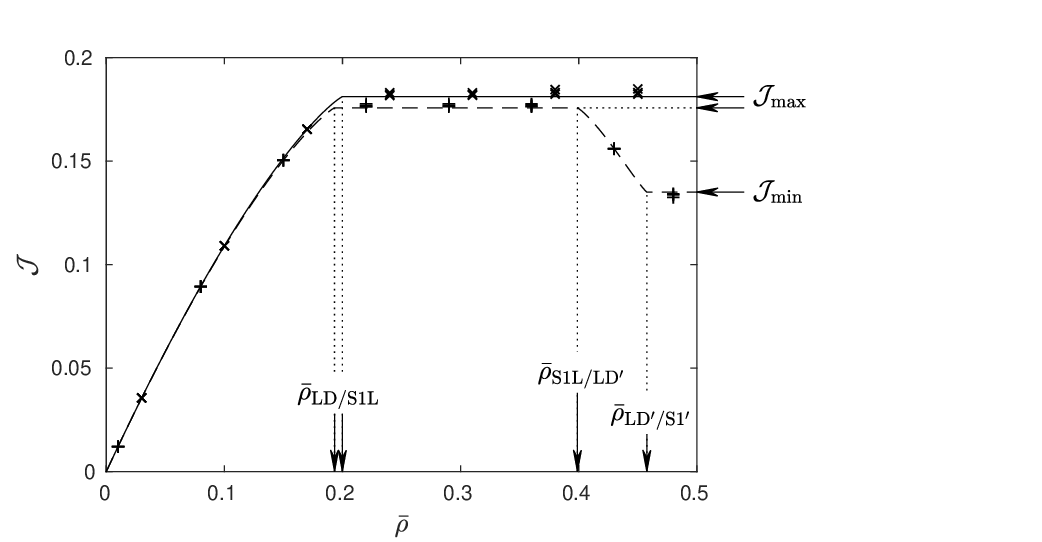}}
	\caption
	{
		Current as a function of the mean density (fundamental diagram) at ${\lambda_{\max} = 1.5}$, ${\lambda_{\min} = 1}$, and ${{v} = 0.15}$ (solid line, ``$\times$''~symbols) or ${{v} = 0.10}$ (dashed line, ``$+$''~symbols). 
		Lines and symbols respectively denote the continuum PA theory and KMC simulations (${{L}=2000,5000,10000}$). 
		Tags are explained in the text. 
	}
	\label{fig:diagramma_rhomedia_corrente}
\end{figure}

Further plausible evidence of exactness for the PA theory is obtained by analyzing the so-called \emph{fundamental diagram} \cite{SchadschneiderChowdhuryNishinari2011}, i.e.~the current $\Jcal$ as a function of the mean density $\bar{\rho}$.
In the framework of the theory, it is convenient to compute the latter as a function of the former, by integrating the appropriate analytical density profiles, evaluated at a given current $\Jcal$.
In figure~\ref{fig:diagramma_rhomedia_corrente} we report the results, for the previously considered values of the parameters (representing the MS and SS regimes), and again we make a comparison with the corresponding simulations (due to the mirror symmetry, we only consider one half of the diagram, for ${\bar{\rho} \in [0,1/2]}$).  
We note that the increasing part of the theoretical curves corresponds to the LD phase, while the decreasing part corresponds to the LD' phase. 
The former occurs in both regimes, for ${\Jcal \in [0,\Jcal_{\max}]}$, whereas the latter occurs only in the SS regime, for ${\Jcal \in [\Jcal_{\min},\Jcal_{\max}]}$. 
In the two cases, the mean density as a function of the current is respectively obtained as 
\begin{subequations} 
	\label{eq:density-vs-current}
	\begin{eqnarray}
	\bar{\rho}_{\mathrm{LD}}(\Jcal)  & = \int_{0}^{1} \left. {\rho}_{1}^{-}(x) \right|_{\Jcal} \rmd x
	\, , \\
	\bar{\rho}_{\mathrm{LD'}}(\Jcal) & = \int_{0}^{1} \left. {\rho}_{2}^{-}(x) \right|_{\Jcal} \rmd x
	\, .
	\end{eqnarray} 
\end{subequations} 
Conversely, the plateau regions, where the current is independent of the mean density, correspond to the various shock phases (at maximal or minimal current). 
By the way we note that, evaluating \eqref{eq:density-vs-current} at $\Jcal_{\max}$ or possibly $\Jcal_{\min}$, we recover the mean-density values at the transitions with the shock phases, previously calculated by \eqref{eq:transizioni_SS}, i.e.
\begin{subequations} 
\begin{eqnarray}
	\bar{\rho}_{\mathrm{LD}}(\Jcal_{\max})  & = \bar{\rho}_{\mathrm{LD/S1L}}
	\, , \\
	\bar{\rho}_{\mathrm{LD'}}(\Jcal_{\max}) & = \bar{\rho}_{\mathrm{S1L/LD'}}
	\, , \\
	\bar{\rho}_{\mathrm{LD'}}(\Jcal_{\min}) & = \bar{\rho}_{\mathrm{LD'/S1'}}
	\, .
\end{eqnarray} 
\end{subequations} 
In figure~\ref{fig:diagramma_rhomedia_corrente} these values are explicitly tagged. 
Regarding the simulations, we have studied just a few values of the mean density in a very accurate way. 
It can be seen that in the smooth phases the agreement is excellent already at the smallest size considered (${L=2000}$), such that the deviation with respect to the theoretical prediction is not even detectable at the scale of the figure.
In the shock phases there is a small residual discrepancy, which however shows a clear tendency to reduce for increasing size. 

\begin{figure}
	\flushright{\includegraphics*[width=0.85\textwidth]{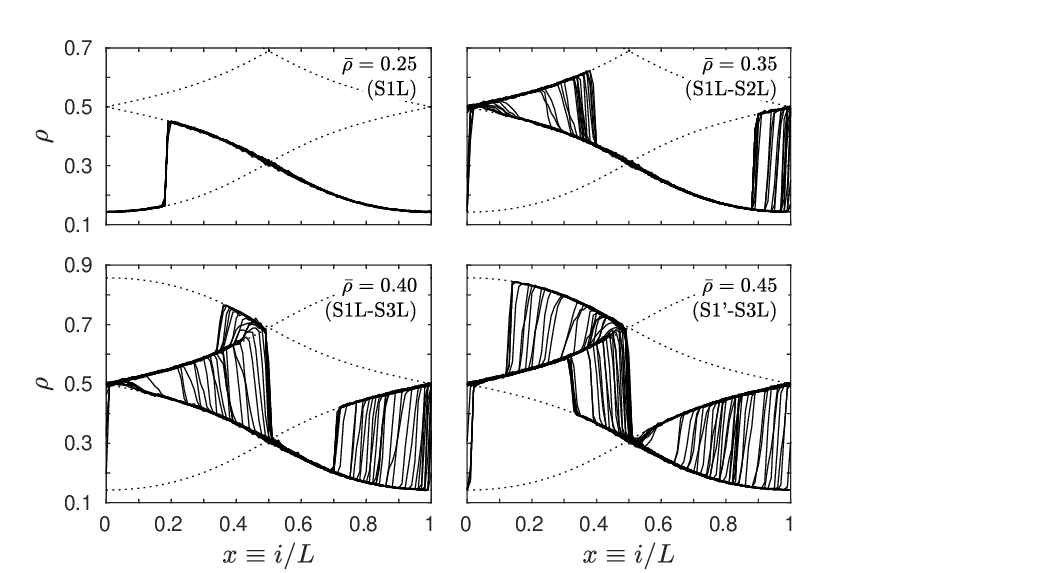}}
	\caption
	{
		Crossover regime for ${\lambda_{\max}/\lambda_{\min} = 1.5}$ and different mean-density values: ${\bar{\rho}=0.25}$ (S1L phase), $0.35$ (S1L-S2L crossover), $0.40$ (S1L-S3L crossover), $0.45$ (S1'-S3L crossover). 
		Solid lines denote KMC simulation results for ${L=10000}$ and ${{v} = 0.1405}$, $0.1410$, $0.1415$, $0.1420$ (only for ${\bar{\rho}=0.40,0.45}$), $0.1425$ (only for ${\bar{\rho}=0.45}$), averaging over different time windows (see the text). 
		Dotted lines denote all solutions of \eqref{eq:profili_analitici} at ${{v} \approx 0.1392}$ (theoretical crossover value) and maximal current ${\Jcal_{\max} = \lambda_{\min} {J}_*}$. 
	}
	\label{fig:profili_multishock_critici}
\end{figure}

In the remainder of this section, we try to give a somewhat more detailed description of what happens at the transition between the MS and SS regimes, which we previously called the crossover regime. 
From the viewpoint of the continuum limit (in the PA theory), we have already noticed that in this situation the two so-called critical points ${x_\circ}$ and ${1-x_\circ}$, occurring in the MS regime, degenerate into a single ``coalescence point'' (in the graphs we see this point split at the left and right edges: ${x=0,1}$). 
As a result, the shock placed at the coalescence point in the S2L and S3L phases is no longer locked, which gives rise to an extra degree of freedom in the shock positions. 
It can therefore be expected that the S2L and S3L phases in the crossover regime are characterized by 2 shocks whose positions vary over time with a random motion (plausibly with a slower kinetic than that of particle hopping), but with a constraint on the relative position imposed by equation \eqref{eq:densita_media} (that is -- physically speaking -- by particle conservation). 
As already mentioned above, we speak in this case of delocalized and synchronized shocks \cite{BanerjeeBasu2020}. 
In order to visualize this phenomenon, we carry out KMC simulations in which, after waiting the usual settling time, we average over a certain number of (disjoint) successive time windows of duration ${t_\mathrm{ave}=10^5}$. 
Figure~\ref{fig:profili_multishock_critici} shows some results, corresponding to the usual rate modulation ratio value ${\lambda_{\max}/\lambda_{\min} = 1.5}$ (so one can still refer to the phase diagram in figure~\ref{fig:phase_diagram_v_rhomedia}). 
The expected phenomenon is clearly visible, that is, by superimposing density profiles computed on different time windows, shocks are observed in different positions. 
In this regard, it should be noted that the choice of time windows shorter than usual does not only have the role of limiting the required computational effort, but also that of better highlighting the shock dynamics itself (too short a window would result in a noisy profile, whereas a very long window would show a smooth profile, without shocks).
Of course, the synchronization effect cannot be appreciated from the figure, but it can actually be verified, albeit quite roughly, by examining the different profiles separately.

Let us now describe in detail the four cases displayed in figure~\ref{fig:profili_multishock_critici}.
At low mean-density values (specifically at ${\bar{\rho}=0.25}$), it can be seen that the density profile never reaches the coalescence point, so the system is completely unaffected by crossover. 
The profile is characterized by a single very stable shock, qualitatively equivalent to that of the S1L phase (see figures \ref{fig:profili_multishock_simulati} and \ref{fig:profili_smallshock_simulati}).
Actually, in the phase diagram of figure~\ref{fig:phase_diagram_v_rhomedia}, the point corresponding to this case is located inside the S1L phase region.
At higher mean-density values (specifically at ${\bar{\rho}=0.35}$), we begin to observe 2 shocks (which for simplicity we will call ``upper'' and ``lower'', relating to density values) and the delocalization phenomenon, that is, shock positions vary, within a certain range, depending on the time window considered.
At one end of the range, the lower shock is positioned at the coalescence point (${x=0,1}$) and the upper one at some position ${x \in (0,1/2)}$. 
The resulting profile is therefore completely analogous to that of the S2L phase (see figure~\ref{fig:profili_multishock_simulati}).
However, as previously mentioned, in this crossover situation the lower shock is no longer locked, i.e.~it can move backwards (in the figure we see it reentering from the opposite side, at ${x=1}$), up to an extreme position ${x \in (1/2,1)}$, which corresponds to the upper shock getting back to the coalescence point (${x=0}$), with vanishing amplitude.
This opposite extreme situation is therefore equivalent to the S1L phase.
In the phase diagram in figure~\ref{fig:phase_diagram_v_rhomedia}, the corresponding point is located along the S1L-S2L section of the crossover line. 
At even higher mean-density values (${\bar{\rho}=0.40,0.45}$) one can observe up to 3 shocks (say ``upper'', ``intermediate'', and ``lower''). 
For both cases, one of the extremes of the delocalization range corresponds to the lower shock positioned at the coalescence point, with an overall profile equivalent to that of the S3L phase (see figure~\ref{fig:profili_multishock_simulati}).
As already seen above, however, the lower shock can move backwards, and this induces (due to the usual conservation constraint) a forward displacement of the upper shock, while, at least initially, the intermediate-shock position remains locked at ${x =1/2}$.
If the displacement of the lower shock goes beyond a certain limit, the upper shock can come to vanish at the ${x=1/2}$ position, after which the intermediate shock gets ``unlocked'' and can proceed backwards.
In this situation the profile is equivalent to that of the S2L phase.
Now, if the mean density is not too high (e.g. ${\bar{\rho}=0.40}$), the intermediate shock can even reach the coalescence point (${x=0}$), with vanishing amplitude, when the lower shock is still in a position ${x \in (1/2,1)}$, so that the extremal profile is still that of the S1L phase (in the diagram of figure~\ref{fig:phase_diagram_v_rhomedia} we are along the S1L-S3L section of the crossover line). 
If, on the other hand, the mean density exceeds a certain limit (e.g. ${\bar{\rho}=0.45}$), it can be seen that the intermediate shock can never reach the coalescence point, but at most a position ${x \in (0,1/2)}$, with nonzero amplitude. 
This corresponds to the fact that the lower shock proceeds backward until it vanishes at position ${x=1/2}$, so that the resulting extremal profile is that of the S1' phase (see figure~\ref{fig:profili_smallshock_simulati}). 
Still with reference to figure~\ref{fig:phase_diagram_v_rhomedia}, in this last case we are indeed along the S1'-S3L section of the crossover line.
In general, we can say that the situations illustrated in figure~\ref{fig:profili_multishock_critici} exhaust the types of delocalized shocks, that can be observed in this model. 
Another general fact, which is worth noting, is that the whole crossover phenomenology, described above, is observed in the simulations over a (narrow) interval of interaction parameter values, not at a single special value, as predicted by the theory in the continuum limit. 
Furthermore, the interval does not even cover the theoretical crossover value (${{v} \approx 0.1392}$), determined by equation \eqref{eq:transizione_MS-SS}. 
In fact, the simulations carried out at the theoretical $v$ value still exhibit extremely stable shocks, qualitatively indistinguishable from those of the SS regime, shown in figure~\ref{fig:profili_smallshock_simulati}.
We have reason to believe that this discrepancy too, like others observed previously, is a finite-size effect, since it can be verified that both the amplitude of the aforementioned interval and the distance from the theoretical value, which are anyway small, also show a clear decreasing tendency, upon increasing size.

\section{Conclusions} 
\label{sec:conclusions}

In this paper we have studied a TASEP with periodic boundary conditions, hopping rates characterized by a smooth spatial modulation (with a unique global minimum), and a NN repulsive interaction. 
The model is in principle analogous to that considered in a previous study \cite{PalGupta2021}, yet with the interaction parameters characterized by a different constraint, denoted as the KLS condition. 
Such a constraint makes the model more amenable to analytical treatment, specifically by means of the so-called PA theory (a generalized mean-field theory, based on a NN pair cluster). 
Actually, we conjecture that, in combination with the KLS condition, the analytical theory is asymptotically exact, except in the vicinity of shocks. 
Although we cannot rigorously prove this statement, we provide considerable numerical evidences. 

Our main contribution is to have elucidated the nature of so-called S~phases, namely, certain non-equilibrium steady states, that the cited previous study \cite{PalGupta2021} had detected as qualitatively distinguished by the more ``universal'' smooth and shock phases \cite{BanerjeeBasu2020}. 
In particular, we have pointed out that such phases exhibit different features, depending on whether one considers intermediate or strong repulsion regimes. 
In the intermediate regime, it turns out that the S~phases can be regarded as subphases of the maximal-current phase, and can also be distinguished into several further subphases, with density profiles displaying up to 3 shocks (MS regime). 
On the other hand, in the strong regime it turns out that the S~phases can be either maximal- or minimal-current phases, but always displaying just 1 shock (SS regime).  
Furthermore, we have pointed out the existence of a crossover regime, emerging in between the aforementioned regimes 
and featuring delocalized and syncronized shocks \cite{BanerjeeBasu2020}. 
The analytical PA theory has allowed us to work out the whole phase diagram of the model in great detail. 
As a consequence of the above conjecture, we claim that such phase diagram may be exact as well, under the KLS assumption, just qualitatively correct otherwise. 
In fact, we have argued that the whole, considerably rich phenomenology emerging in this model can be traced back to the onset of a bimodal current-density relation, so that the choice of the KLS condition is relevant only at a quantitative level. 
Thus, we claim that our results can describe in a qualitatively correct way the physics of a whole class of similar models, differing only by the absence of the KLS condition, and including in particular the one studied in \cite{PalGupta2021}. 
In the course of our work we have indeed collected several evidences of such a claim, that have not been reported. 
\begin{figure}
	\flushright{\includegraphics*[width=0.85\textwidth]{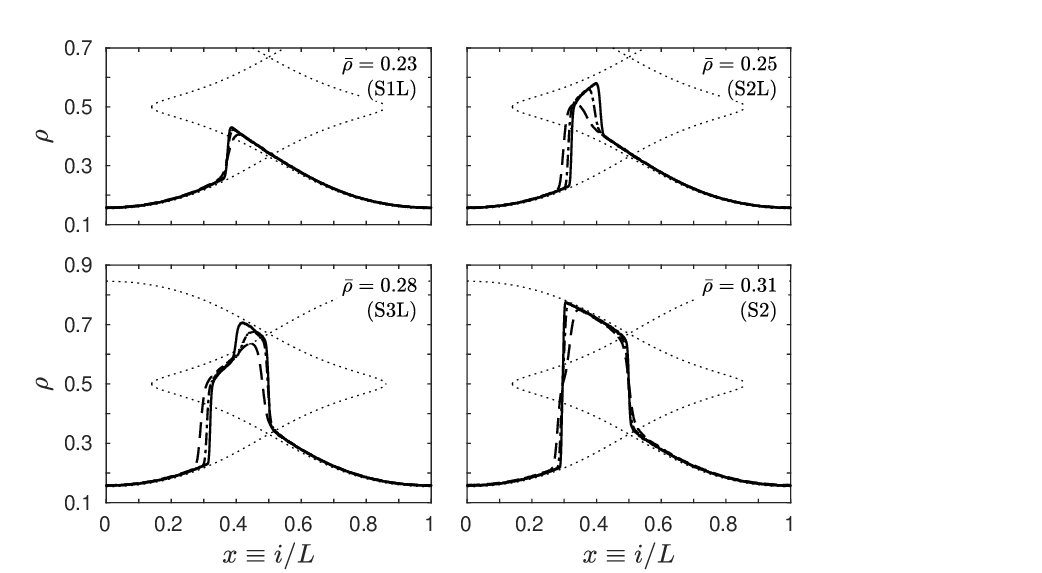}}
	\caption
	{
		MS regime in the DME assumption (see the text), for ${\lambda_{\max}/\lambda_{\min} = 1.5}$, ${{v} = 0.09}$, and different mean-density values: ${\bar{\rho}=0.23}$ (S1L phase), $0.25$ (S2L phase), $0.28$ (S3L phase), $0.31$ (S2 phase). 
		Thick lines denote KMC simulation results for ${L=2000}$ (dashed lines), $5000$ (dash-dotted lines), $10000$ (solid lines). 
		Thin dotted lines denote all possible solutions of \eqref{eq:profili_analitici} at maximal current ${\Jcal_{\max} = \lambda_{\min} {J}_*}$. 
	}
	\label{fig:profili_multishock_simulati_DME}
\end{figure}
Here, in figure~\ref{fig:profili_multishock_simulati_DME}, we report just one significant example, relating to the analogous model with DME condition (i.e.~the one considered in \cite{PalGupta2021}), in the intermediate-repulsion (MS) regime. 
We can see in particular that the PA theory quantitatively fails, especially in predicting the location of the critical points. 
However, the sequence of density profiles, upon incresing the mean particle density, turns out to be qualitatively equivalent to that obtained with the KLS condition in the same regime (figure \ref{fig:profili_multishock_simulati}).  
 
Let us finally note that the present work naturally opens the way to the investigation of a number of related models, for which the analytical theory developed here could be applied without relevant changes. 
Among such extensions, which are of course beyond the scope of this paper, we first have in mind the very same model with open boundary conditions (\cite{GoswamiChatterjeeMukherjee2022} considers a closely related model, though without NN interaction). 
We also include the possibility of relaxing some of the restrictive assumptions about hopping rates, in particular \eqref{eq:gamma} and \eqref{eq:simmetria_parametri}. 
Relaxing \eqref{eq:gamma} means to consider that the interaction itself may feature a spatial modulation, whereas relaxing \eqref{eq:simmetria_parametri} amounts to remove the mirror symmetry of the current-density relation. 
The former issue is mainly methodological (specifically to test whether the PA theory is still reliable in such a case), the latter is meant to make the model slightly more realistic as a model for vehicular traffic \cite{SchadschneiderChowdhuryNishinari2011,AntalSchutz2000}. 
It would also be interesting to carry out a deeper investigation about the random motion of shocks in the crossover regime, possibly developing an ad-hoc theory based on a Fokker-Planck equation, along the lines of \cite{BanerjeeBasu2020} and \cite{ReichenbachFranoschFrey2008}. 
Furthermore, one could investigate to what extent the observed phenomena can be affected by any concurrent processes, such as Langmuir kinetics \cite{ParmeggianiFranoschFrey2004,PierobonMobiliaKouyosFrey2006} and/or local stochastic resetting \cite{MironReuveni2021,PelizzolaPrettiZamparo2021,PelizzolaPretti2022,NagarGupta2023}.
We are considering these types of questions as possible topics for future work. 

\appendix

\section{PA theory in the continuum limit}
\label{app:cont_lim}

In this appendix we describe the continuum (or hydrodynamic) limit of the PA theory, in order to determine the current-density relation of our system. 
As we consider the steady state, we drop the time index in all equations, and set at zero the time derivatives in \eqref{eq:continuity} and \eqref{eq:phipunto}. 
From \eqref{eq:continuity} we obviously obtain ${\Jcal_{i} = \Jcal}$ for all $i$ (particles are conserved), so that equations \eqref{eq:corrente} and \eqref{eq:phipunto} respectively become  
\begin{eqnarray}
	\Jcal & = \Jcal_{i}(0,0) + \Jcal_{i}(0,1) + \Jcal_{i}(1,0) + \Jcal_{i}(1,1)
	\label{eq:cont_lim:corrente1}
	\, , \\
	0 & = \Jcal_{i-1}(0,1) + \Jcal_{i-1}(1,1) - \Jcal_{i+1}(1,0) - \Jcal_{i+1}(1,1)
	\label{eq:cont_lim:correlaz1}
	\, . 
\end{eqnarray}
In the continuum limit, one assumes that marginal probabilities depend on the node index $i$, only through the scaled position variable ${x \equiv i/L}$, as well as hopping rates do by construction. 
In the above equations we neglect the terms that vanish in the ${L \to \infty}$ limit, which amounts to discarding finite differences in the position index. 
As a consequence, taking into account \eqref{eq:corrente_elem}, \eqref{eq:gamma} and \eqref{eq:reduced_current}, we obtain 
\begin{eqnarray}
	{J} & = {q} \, {P}[0101] + {r} \, {P}[1100] + {p} \, {P}[0100] + {s} \, {P}[1101] 
	\label{eq:cont_lim:corrente2}
	\, , \\
	0   & = {q} \, {P}[0101] - {r} \, {P}[1100] 
	\label{eq:cont_lim:correlaz2}
	\, , 
\end{eqnarray}
where all space-dependent quantities are evaluated at the same scaled position $x$ (which for simplicity we no longer display). 
Note that the resulting equations are formally equivalent to those derived for a uniform system, for instance in \cite{MidhaKolomeiskyGupta2018jstat}, but here we understand that the probabilities ${P}[k10n]$ and the reduced current ${J}$ depend on $x$.\footnote{In principle we could assume that the parameters ${p},{q},{r},{s}$ also depend (smoothly) on $x$, without invalidating the derivation. However, in the article we do not analyze cases of this type.} 
Plugging equation \eqref{eq:cont_lim:correlaz2} into \eqref{eq:cont_lim:corrente2}, the latter simplifies to 
\begin{equation}
	{J} = 2{q} \, {P}[0101] + {p} \, {P}[0100] + {s} \, {P}[1101] 
	\, . 
	\label{eq:cont_lim:corrente3}
\end{equation}
We then introduce the PA \eqref{eq:pairwapp} (where we still neglect position index differences) into \eqref{eq:cont_lim:corrente3} and \eqref{eq:cont_lim:correlaz2}, yielding respectively 
\begin{eqnarray}
	{J} = \frac{{P}[10] \, {P}[01]}{{P}[1] \, {P}[0]} 
	\left( \vphantom{{}^{2}} 2{q} \, {P}[01] + {p} \, {P}[00] + {s} \, {P}[11] \right)
	\label{eq:cont_lim:corrente4}
	\, , \\ 
	{P}[11] \, {P}[00] = \frac{\,{q}\,}{{r}} \, {{P}[01]}^{2}
	\, . 
	\label{eq:cont_lim:correlaz4}
\end{eqnarray}
The probabilities can be parameterized in terms of local densities and NN correlations, as done in section~\ref{sec:methods}. 
According to \eqref{eq:site_marginal}, we have of course 
\begin{equation} 
	{P}[1] = \rho 
	\, , \qquad 
	{P}[0] = 1 - \rho
	\, . 
	\label{eq:cont_lim:site_marginal}
\end{equation} 
On the other hand, following \cite{PelizzolaPrettiPuccioni2019}, it is convenient to parameterize pair probabilities by means of a particular \emph{correlator} variable, defined as 
\begin{equation}
	\eta \triangleq \frac{{P}[10]}{{P}[1] \, {P}[0]}
	\, .  
	\label{eq:cont_lim:definizione_eta}
\end{equation}
Defining also the quantity 
\begin{equation}
	{I} \triangleq \rho \, (1-\rho) 
	\, 
	\label{eq:cont_lim:definizione_I}
\end{equation}
(which we shall call \emph{standard current}, in that it represents the current-density relation of ``ordinary'' TASEP with unit hopping rate), according to \eqref{eq:pair_marginal} we have 
\begin{subequations} 
	\label{eq:cont_lim:pair_marginal}
	\begin{eqnarray}
		{P}[10] & = {P}[01] = \eta {I}
		\, , \\
		{P}[00] & = 1-\rho - \eta {I}
		\, , \\
		{P}[11] & = \rho - \eta {I}
		\, . 
	\end{eqnarray}
\end{subequations} 
Now, using equations \eqref{eq:cont_lim:site_marginal}-\eqref{eq:cont_lim:pair_marginal}, we can rewrite \eqref{eq:cont_lim:corrente4} and \eqref{eq:cont_lim:correlaz4} respectively as 
\begin{eqnarray}
	{J} = \eta^2 {I} \left[ \vphantom{{}^{2}} (2{q}-{p}-{s}) \, \eta {I} + {p} - ({p}-{s}) \rho \right]
	\label{eq:cont_lim:corrente5}
	\, , \\
	\eta^2 {I} = \frac{\eta-1}{1-{q}/{r}} 
	\label{eq:cont_lim:correlaz5}
    \, .
\end{eqnarray}
Let us further observe that a repeated use of \eqref{eq:cont_lim:correlaz5} allows us to derive a simpler expression for the current, linear in $\eta$. 
Actually, plugging \eqref{eq:cont_lim:correlaz5} first into \eqref{eq:cont_lim:corrente5} and then into the resulting expression, and defining the combinations of rates below 
\begin{subequations} 
\label{eq:cont_lim:definizioni_rtilde_a_b}
\begin{eqnarray}
	\tilde{r} & \triangleq {r} \ \frac{{p}+{s}-2{q}}{{r}-{q}} 
	\label{eq:cont_lim:definizione_rtilde}
	\, , \\
	{a} & \triangleq \frac{{r}}{{r}-{q}} - \frac{{p}}{{p}+{s}-2{q}}
	\label{eq:cont_lim:definizione_a}
	\, , \\
	{b} & \triangleq \frac{{p}-{s}}{{p}+{s}-2{q}} 
	\label{eq:cont_lim:definizione_b}
	\, , 
\end{eqnarray}
\end{subequations} 
we arrive at 
\begin{eqnarray}
	{J} 
	& = \tilde{r} \left[ \vphantom{{}^{2}} \eta {I} - (\eta-1) ({a} + {b} \rho)  \right]
	\, . 
	\label{eq:cont_lim:corrente6}
\end{eqnarray}
We stress the fact that equation \eqref{eq:cont_lim:correlaz5} establishes a relation between $\eta$ and ${I}$, whereas ${I}$ directly depends on $\rho$ through \eqref{eq:cont_lim:definizione_I}. 
As a consequence, the right-hand side of \eqref{eq:cont_lim:corrente6}, in combination with \eqref{eq:cont_lim:correlaz5} and \eqref{eq:cont_lim:definizione_I}, completely specifies the current-density function ${F}_{{p},{q},{r},{s}}(\rho)$, introduced in section \ref{sec:methods}, where the density and the rates respectively play the roles of independent variable and parameters. 
The property of homogeneity with respect to the parameters, stated by equation~\eqref{eq:omogeneita}, can easily be verified. 

Let us also note that \eqref{eq:cont_lim:corrente6} is formally equivalent to equation (34) in reference \cite{PelizzolaPrettiPuccioni2019}, which determines the steady-state current of a similar model, with hopping rates depending on forward and backward occupation numbers, but (i) independent of position and (ii) satisfying the KLS condition. 
At odds with the cited paper, here we have different (more general) expressions for the $\tilde{r},{a},{b}$ coefficients and we understand that all the variables depend on the scaled position $x$. 
Actually, if in equations \eqref{eq:cont_lim:definizioni_rtilde_a_b} we force the KLS condition \eqref{eq:condizione_KLS}, then we formally recover the coefficients reported in \cite{PelizzolaPrettiPuccioni2019}, equations (35), and in particular ${\tilde{r} = {r}}$.  
Let us also observe that, as mentioned above, according to equation \eqref{eq:cont_lim:correlaz5} the correlator $\eta$ depends on the density $\rho$ only through the standard current ${I}$, which is obviously symmetric under the ``particle-hole'' transformation ${\rho \mapsto (1-\rho)}$. 
As a consequence, assuming \eqref{eq:simmetria_parametri}, that is ${{b}=0}$, induces the same symmetry in the current-density relation. 
Except the general expressions above, all results reported in this paper are obtained by taking the symmetry hypothesis \eqref{eq:simmetria_parametri}. 
Plugging the latter into \eqref{eq:cont_lim:definizione_rtilde} and \eqref{eq:cont_lim:definizione_a}, along with the interaction parameter definition \eqref{eq:definizione_v}, we obtain expressions for the remaining coefficients, $\tilde{r}$ and ${a}$, as functions of the basic independent parameters ${v}$ and ${q}$, namely 
\begin{subequations} 
\label{eq:cont_lim:definizioni_rtilde_a_semplici}
\begin{eqnarray}
	\tilde{r} & = \frac{2(1-{q})}{1-{v}^2} 
	\label{eq:cont_lim:definizione_rtilde_semplice}
	\, , \\
	{a} & = \frac{1}{1-{v}^2} - \frac{1}{2(1-{q})}
	\label{eq:cont_lim:definizione_a_semplice}
	\, . 
\end{eqnarray}
\end{subequations} 

\begin{figure}
	\flushright{\includegraphics*[width=0.85\textwidth]{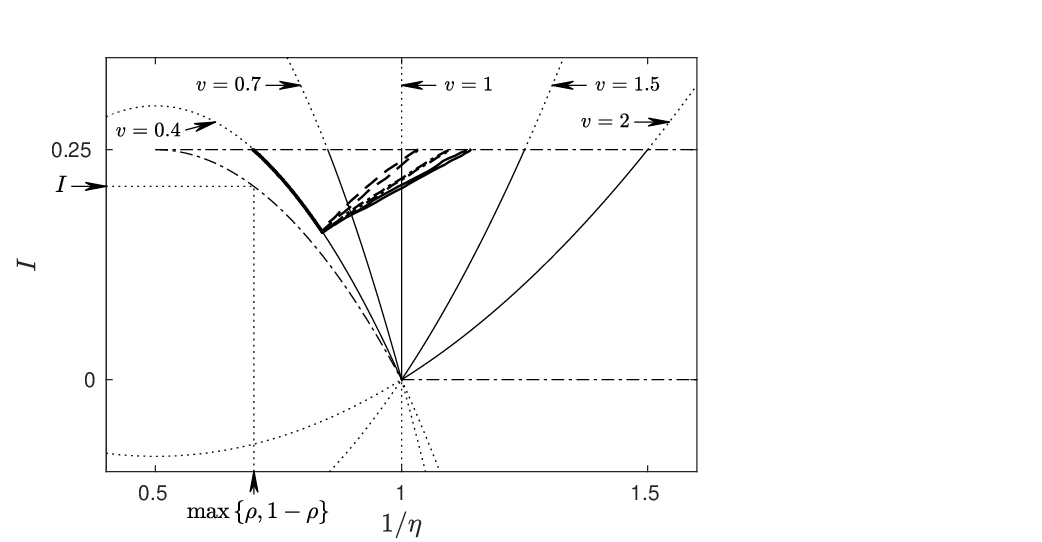}}
	\caption{
		The admissible region in the plane ${I}$ vs $1/\eta$ is delimited by a thin dash-dotted line. 
		Thin solid lines represent the correlator-density relation \eqref{eq:cont_lim:correlaz6}, for different values of ${v}$ (the analytical continuations outside the admissible region are denoted by dotted lines). 
		Thicker lines denote KMC simulation results for ${{v}=0.4}$, ${\lambda_{\max}/\lambda_{\min}=1.5}$, ${\bar{\rho}=0.5}$ and ${L=2000}$ (dashed line), ${5000}$ (dash-dotted line), ${10000}$ (solid line).
	}
	\label{fig:piano_unosueta_I}
\end{figure}

In the remainder of this section we report some details about the dependence of $\eta$ on ${I}$, stated by \eqref{eq:cont_lim:correlaz5}, which we need to explicitly determine the current-density relation, and which we shall denote (with a slight abuse of language) as \emph{correlator-density relation}. 
We stress the fact that this relation is unaffected by the symmetry assumption \eqref{eq:simmetria_parametri}. 
Let us first characterize the ranges of admissible values for ${I}$ and $\eta$. 
In fact, only inequality constraints are needed, in order to ensure that the probabilities \eqref{eq:cont_lim:site_marginal} and \eqref{eq:cont_lim:pair_marginal} stay between $0$ and $1$ (no equality constraint is required, since the parameterizations intrinsically satisfy normalization and compatibility conditions\footnote{By compatibility we mean that 1-node distributions are actually marginals of 2-node distributions.}). 
As far as \eqref{eq:cont_lim:site_marginal} are concerned (single-node probabilities), we obviously require ${0 \le \rho \le 1}$, which entails ${0 \le {I} \le 1/4}$. 
Moreover, regarding \eqref{eq:cont_lim:pair_marginal} (pair probabilities), the required condition turns out to be 
\begin{equation} 
	\frac{\,1\,}{\eta} \ge \max \{ \rho, 1-\rho \}
	= \frac{1 + \sqrt{1 - 4{I}\vphantom{(1-{v}^2)}}}{2}
	\, .
	\label{eq:cont_lim:disuguaglianza_unosueta_I}
\end{equation} 
The latter equality follows from \eqref{eq:cont_lim:definizione_I}, which states that a given (admissible) standard-current value has two possible corresponding densities, with unit sum, precisely 
\begin{equation}
	\rho = \frac{1 \pm \sqrt{1 - 4{I}\vphantom{(1-{v}^2)}}}{2} 
	\, .
	\label{eq:cont_lim:rho_vs_I}
\end{equation} 
Now, in terms of the reciprocal variable $1/\eta$, the correlator-density relation \eqref{eq:cont_lim:correlaz5} reads 
\begin{equation} 
	(1-{v}^2) {I} 
	= \frac{\,1\,}{\eta} \left( 1-\frac{\,1\,}{\eta} \right) 
	\, , 
	\label{eq:cont_lim:correlaz6}
\end{equation} 
clearly analogous to \eqref{eq:cont_lim:definizione_I}. 
Consequently, also this equation has in principle two solutions for $1/\eta$, with unit sum, but we easily argue that only the larger one 
\begin{equation}
	\frac{\,1\,}{\eta} = \frac{1 + \sqrt{1 - 4(1-{v}^2){I}}}{2} 
	\, 
	\label{eq:cont_lim:correlaz7}
\end{equation}
satisfies \eqref{eq:cont_lim:disuguaglianza_unosueta_I}, with equality occurring precisely for ${{v}=0}$. 
For more clarity, in figure~\ref{fig:piano_unosueta_I} we display the admissible region in the plane ${I}$ vs $1/\eta$, along with some instances of the correlator-density relation \eqref{eq:cont_lim:correlaz6}, obtained for different values of ${v}$. 
We note in particular that, excluding the borderline case ${{I} = 0}$, ${\eta = 1}$, the regions corresponding to ${{v} < 1}$ or ${{v} > 1}$ are respectively characterized by ${\eta > 1}$ or ${\eta < 1}$. 
Remembering the correlator definition \eqref{eq:cont_lim:definizione_eta}, this physically means that we have, respectively, a larger or smaller probability to find an empty node, if we know that one of its neighbors is occupied. 
These are clearly the fingerprints of a repulsive or attractive interaction, respectively. 
Some results from KMC simulations, reported as well in figure~\ref{fig:piano_unosueta_I}, confirm that the PA theory in the continuum limit provides extremely accurate (possibly exact) results in regions characterized by smooth profiles. 
On the other hand, a definite deviation corresponds to a shock region, where as usual the results are highly size-dependent.  
By the way, we also note that in the shock region the correlator $\eta$ can become smaller than 1, even in case of repulsive interaction, and that the effect is more pronounced for larger system size.

\section{Transition to bimodality}
\label{app:ph_trans}

Making use of the continuum PA theory, in this appendix we determine the transition where, as described in section~\ref{sec:methods}, the current-density relation becomes bimodal, with the central minimum at density ${\rho_\circ = 1/2}$ and the two side maxima at densities ${\rho_* = 1/2 \pm \Delta}$. 
We understand that this calculation refers to a ``particle-hole-symmetric'' case, where \eqref{eq:simmetria_parametri} holds, and the appropriate expression for the (reduced) current is \eqref{eq:cont_lim:corrente6} with ${{b} = 0}$, namely 
\begin{equation}
	{J} = \tilde{r} \left[ \vphantom{{}_{|}} \eta {I} - {a} \, (\eta - 1) \right]
	\, . 
	\label{eq:ph_trans:corrente1}
\end{equation}
According to the correlator-density relation \eqref{eq:cont_lim:correlaz6}, we can write 
\begin{equation}
	\eta {I} = \frac{1 - 1/\eta}{1-{v}^2} 
	\, . 
	\label{eq:ph_trans:correlaz1}
\end{equation}
Moreover, taking into account the parameter expressions \eqref{eq:cont_lim:definizioni_rtilde_a_semplici}, we have 
\begin{equation}
	{a} = \frac{1 - 1/\tilde{r}}{1-{v}^2}
	\, . 
	\label{eq:ph_trans:espressione_a}
\end{equation}
Plugging \eqref{eq:ph_trans:correlaz1} and \eqref{eq:ph_trans:espressione_a} into \eqref{eq:ph_trans:corrente1}, we obtain 
\begin{equation}
	{J} = \frac{ \tilde{r} \, (1 - 1/\eta) - (\tilde{r} - 1) \, (\eta - 1) }{ 1-{v}^2 }
	\, , 
	\label{eq:ph_trans:J_di_eta}
\end{equation}
where we see that, apart from parameters, the current is expressed as a function of the only correlator variable $\eta$. 
We also recall that, as stated by \eqref{eq:cont_lim:correlaz7}, $\eta$ does not depend on $\rho$ explicitly, but only through the standard current ${I}$. 
As a consequence, the derivative can be evaluated as 
\begin{equation}
	\frac{\rmd {J}}{\rmd \rho}
	= 
	\frac{\rmd {J}}{\rmd \eta} \
	\frac{\rmd \eta}{\rmd {I}} \
	\frac{\rmd {I}}{\rmd \rho} \ . 
\end{equation}
We can now make the following observations. 
\begin{enumerate} 
	
	\item As discussed in \ref{app:cont_lim}, the $\eta$~vs~${I}$ relation \eqref{eq:cont_lim:correlaz7} turns out to be strictly monotonic (excluding the trivial case ${{v} = 1}$), so we always have 
	\begin{equation}
		\frac{\rmd \eta}{\rmd {I}} \ne 0
		\, . 
	\end{equation}
	
	\item From the standard-current definition \eqref{eq:cont_lim:definizione_I}, we have 
	\begin{equation}
		\frac{\rmd {I}}{\rmd \rho} = 1 - 2 \rho 
		\, , 
	\end{equation}
	which entails that ${\rho_\circ = 1/2}$ is always a stationary point for the current-density relation. 
	
	\item From equation \eqref{eq:ph_trans:J_di_eta} we immediately get
	\begin{equation}
		\frac{\rmd {J}}{\rmd \eta} = \frac{ \tilde{r} / \eta^2 - (\tilde{r} - 1) }{ 1-{v}^2 }
		\, , 
	\end{equation}
	from which we argue that an extra stationary point may appear for ${\eta = \eta_*}$, where 
	\begin{equation}
		\frac{1}{\,\eta_*} = \sqrt{1 - \frac{\,1\, }{\tilde{r}}} 
		\, . 
		\label{eq:ph_trans:unosuetastar}
	\end{equation}
	Equation \eqref{eq:cont_lim:correlaz6} allows us to obtain the corresponding standard-current value ${I}_*$, which in turn corresponds to 2 possible densities $\rho_*$, determined by \eqref{eq:cont_lim:rho_vs_I}. 
	
\end{enumerate} 

In order for the extra stationary points to be placed in the physically meaningful interval (${0 < \rho_* < 1}$), also being distinct from the ``normal'' stationary point (${\rho_* \ne 1/2}$), the required condition in terms of standard current is, according to \eqref{eq:cont_lim:definizione_I}, ${0 < {I}_* < 1/4}$. 
The latter condition corresponds, through equation \eqref{eq:cont_lim:correlaz7}, to different conditions for the correlator, depending on whether ${{v} < 1}$ or ${{v} > 1}$. 
Altogether, we can write 
\begin{equation}
	\frac{1 + {v}}{2} \lessgtr \frac{1}{\,\eta_*} \lessgtr 1 \qquad \mbox{if} \ {v} \lessgtr 1 
	\, . 
	\label{eq:ph_trans:condizioni_unosuetastar}
\end{equation}
Furthermore, analyzing equation \eqref{eq:ph_trans:unosuetastar}, one can argue that \eqref{eq:ph_trans:condizioni_unosuetastar} maps to respective conditions for $\tilde{r}$, namely 
\begin{equation}
	\tilde{r} \gtrless \frac{4}{(1-{v})(3+{v})} \qquad \mbox{if} \ {v} \lessgtr 1 
	\, . 
	\label{eq:ph_trans:condizioni_rtilde}
\end{equation}
Let us observe that the right-hand side of the above inequalities turns out to be either larger than 1 (for ${{v} < 1}$) or negative (for ${{v} > 1}$). 
As a consequence, in both cases \eqref{eq:ph_trans:condizioni_rtilde} inherently satisfies the realness condition for the correlator expression \eqref{eq:ph_trans:unosuetastar}, which is of course ${1/\tilde{r} < 1}$. 
Now, plugging \eqref{eq:cont_lim:definizione_rtilde_semplice} into \eqref{eq:ph_trans:condizioni_rtilde}, we can rephrase the conditions in terms of the basic parameters ${v}$ and ${q}$, obtaining 
\begin{equation}
	\frac{1-{q}}{(1-{v})(1+{v})} \gtrless \frac{2}{(1-{v})(3+{v})} \qquad \mbox{if} \ {v} \lessgtr 1 
	\, , 
\end{equation}
where we immediately note that the ${(1-{v})}$ term can be simplified. 
Since this term is respectively positive for ${{v} < 1}$ or negative for ${{v} > 1}$, in the latter case (only) the inequality sign gets reversed, so that in the end we have a unique inequality, which can be written as 
\begin{eqnarray}
	{q} < \frac{1-{v}}{3+{v}}
	\, . 
	\label{eq:ph_trans:phase_transition}
\end{eqnarray}
Following the above derivation, one can also argue that \eqref{eq:ph_trans:phase_transition} with an equality sign corresponds to the borderline case ${{I}_* = 1/4}$, ${\rho_* = 1/2}$, which represents, in the current-density relation, a degeneration of the side maxima into the ``normal'' central maximum. 
Such a behavior makes reason to regard this phenomenon as a sort of second-order phase transition. 
Let us finally remark that in principle we have correctly taken into account both possibilities ${{v} < 1}$ or ${{v} > 1}$, but in the end we deduce that, for physically meaningful (i.e.~positive) ${q}$~values, \eqref{eq:ph_trans:phase_transition} can be verified only for ${{v} < 1}$, that is for repulsive interactions. 
By means of \eqref{eq:ph_trans:condizioni_rtilde}, we have previously argued that this case also implies ${\tilde{r} > 1}$. 
This last remark will come in handy in the next section.

\section{Maximal and minimal currents}
\label{app:max_e_min}

Still in the framework of the continuum PA theory, in this appendix we consider the stationary points of the current-density relation (in the particle-hole-symmetric case), and determine the corresponding values of the reduced current. 

We first consider the normal stationary point, which, as we have seen before, is always present, and placed at a density value ${\rho_\circ = 1/2}$. 
Let us recall that this point may change from being a maximum to a (local) minimum, when the current-density relation undergoes the transition to bimodality, as described in section~\ref{sec:methods}. 
According to \eqref{eq:cont_lim:definizione_I}, the corresponding standard current is of course ${{I}_\circ = 1/4}$, and hence from \eqref{eq:cont_lim:correlaz7} the corresponding (reciprocal) correlator turns out to be  
\begin{eqnarray}
	\frac{1}{\,\eta_\circ} = \frac{1+{v}}{2}
	\, . 
	\label{eq:max_e_min:unosuetazero}
\end{eqnarray}
To evaluate the current, we can at first plug the above expression into \eqref{eq:ph_trans:J_di_eta}, finding
\begin{equation}
	{J}_\circ 
	= \frac{ \tilde{r} \, (1 - 1/\eta_\circ) - (\tilde{r} - 1) \, (\eta_\circ - 1) }{ 1-{v}^2 } 
	= \frac{1 - (1 - {v}) \, \tilde{r}/2}{{(1 + {v})}^{2}}
	\, . 
\end{equation}
Then, taking into account the expression of $\tilde{r}$ as a function of the basic parameters ${v}$ and ${q}$, namely \eqref{eq:cont_lim:definizione_rtilde_semplice}, we can finally obtain 
\begin{equation}
	{J}_\circ = \frac{\hphantom{(} {q} + {v} \hphantom{)^{3}}}{{(1 + {v})}^{3}}
	\, . 
	\label{eq:max_e_min:Jzero}
\end{equation}

Next we consider the extra stationary points (side maxima) appearing at the transition. 
As discussed in \ref{app:ph_trans}, at such points the correlator takes value $\eta_*$, expressed by equation \eqref{eq:ph_trans:unosuetastar} as a function of $\tilde{r}$. 
To evaluate the corresponding current, we can now plug \eqref{eq:ph_trans:unosuetastar} still into \eqref{eq:ph_trans:J_di_eta}, obtaining 
\begin{equation}
	{J}_* 
	= \frac{ \tilde{r} \, (1 - 1/\eta_*) - (\tilde{r} - 1) \, (\eta_* - 1) }{ 1-{v}^2 } 
	= \frac{ \left( \sqrt{\tilde{r} \vphantom{q}} - \sqrt{\tilde{r}-1 \vphantom{q}} \, \right)^{2} }{ 1-{v}^2 } 
	\label{eq:max_e_min:Jstar_tmp}
	\, . 
\end{equation}
Note that the algebraic steps to obtain the rightmost expression in \eqref{eq:max_e_min:Jstar_tmp} pose no problem of sign (nor of realness of the square roots), because we have previously argued (see \ref{app:ph_trans}) that the occurrence of extra stationary points entails ${{v} < 1}$ and thence ${\tilde{r} > 1}$. 
In the end, again expressing $\tilde{r}$ as a function of ${v}$ and ${q}$ by \eqref{eq:cont_lim:definizione_rtilde_semplice}, we can write
\begin{equation}
	{J}_* = \left( \frac{ \sqrt{\vphantom{{v}^2} 2-2{q}} - \sqrt{1+{v}^2-2{q}} }{ 1-{v}^2 } \, \right)^{2} 
	\, . 
	\label{eq:max_e_min:Jstar}
\end{equation}
For completeness let us finally report, as a function of ${v}$ and ${q}$, also an expression for the corresponding correlator, that is 
\begin{equation}
	\frac{1}{\,\eta_*} 
	= \sqrt{\frac{1+{v}^2-2{q}}{2-2{q}}} 
	\, , 
	\label{eq:max_e_min:unosuetastar}
\end{equation}
obtained of course by plugging \eqref{eq:cont_lim:definizione_rtilde_semplice} into \eqref{eq:ph_trans:unosuetastar}, and an expression for the ratio 
\begin{equation}
	\frac{{J}_*}{{J}_\circ} = \frac{ 1 + {v} }{ {q} + {v} } \left( \frac{ \sqrt{\vphantom{{v}^2} 2-2{q}} - \sqrt{1+{v}^2-2{q}} }{ 1-{v} } \, \right)^{2} 
	\, , 
	\label{eq:max_e_min:JstarsuJzero}	
\end{equation}
immediately obtained from \eqref{eq:max_e_min:Jzero} and \eqref{eq:max_e_min:Jstar}. 
Note that the right-hand side of equation \eqref{eq:max_e_min:JstarsuJzero} can easily be verified to take value $1$ at the bimodality transition \eqref{eq:phase_transition}. 

%
%
%
%
%

%
%
%
%
%

\section{Inverting the current-density relation}
\label{app:dens_prof}

In this last appendix we show how to derive the standard current ${I}$ as a function of the reduced current ${J}$ and of the model parameters, once again in the framework of the continuum PA theory. 
The resulting equation, in combination with \eqref{eq:reduced_current} and \eqref{eq:cont_lim:rho_vs_I}, allows us to determine density profiles analytically, at given values of the physical current $\Jcal$. 
The calculation still refers to a particle-hole-symmetric case, for which the appropriate expression of the current is \eqref{eq:ph_trans:corrente1}, equivalent to 
\begin{equation}
	{J} = \tilde{r} \left[ \vphantom{{}^{2}} {a} - \eta \, ({a} - {I}) \right]
	\, . 
	\label{eq:I_aafo_J:corrente1}
\end{equation}
Assuming to fix $J$, the above equation represents a linear relation between the standard current ${I}$ and the reciprocal correlator $1/\eta$, that is 
\begin{equation}
	\frac{\,1\,}{\eta}  = \frac{{a} - {I}}{{a} - {J}/\tilde{r}} 
	\, . 
\end{equation}
Another equation relating the above two quantities is provided by \eqref{eq:cont_lim:correlaz6}, the so-called correlator-density relation. 
Plugging the former into the latter, we immediately obtain a quadratic equation for ${I}$, whose coefficients depend on the reduced current ${J}$, besides model parameters. 
This equation can be written as 
\begin{equation}
	{I}^{2} - {I} \, \psi({J}) + {J} \, \psi(0) = 0 
	\, ,
	\label{eq:I_aafo_J:equazione_IdatoJ} 
\end{equation}
where we have defined a symbol for the first-order coefficient, explicitly denoting the dependence on ${J}$, that is 
\begin{equation}
	\psi({J}) \triangleq {a} + {J}/\tilde{r} - (1-{v}^2){({a}-{J}/\tilde{r})}^{2} 
	\, . 
	\label{eq:I_aafo_J:definizione_psi}
\end{equation}
Note that in \eqref{eq:I_aafo_J:equazione_IdatoJ} the dependence on the model parameters is fully incorporated into the ${\psi({J})}$ function. 
In particular, the fact that the zeroth-order coefficient can be written as ${{J} \, \psi(0)}$ can easily be argued from \eqref{eq:I_aafo_J:definizione_psi}, taking into account the parameter identity \eqref{eq:ph_trans:espressione_a}. 
For completeness, we can also determine $\psi({J})$ in terms of only the elementary parameters ${v}$ and ${q}$, by plugging into \eqref{eq:I_aafo_J:definizione_psi} the appropriate expressions for $\tilde{r}$ and ${a}$, namely equations \eqref{eq:cont_lim:definizioni_rtilde_a_semplici}. 
After some algebra we obtain 
\begin{equation}
	\psi({J}) 
	= \frac{(1-2{q}+{v}^2) + 2(2-3{q}+{v}^2)(1-{v}^2){J} - {(1-{v}^2)}^{3}{J \vphantom{(1-{v}^2)}}^{2}}{4{(1-{q})}^{2}} 
	\, . 
\end{equation}


\section*{References}

\begin{thebibliography}{99} 
	
	\bibitem{ChouMallickZia2011} Chou~T, Mallick~K and Zia~R~K~P 2011 Non-equilibrium statistical mechanics: from a paradigmatic model to biological transport {\it Rep. Prog. Phys.} {\bf 74} 116601 
	
	\bibitem{SchadschneiderChowdhuryNishinari2011} Schadschneider~A, Chowdhury~D and Nishinari~K 2011 {\it Stochastic Transport in Complex Systems} (Amsterdam: Elsevier)
	
	\bibitem{KriecherbauerKrug2010} Kriecherbauer~T and Krug~J 2010 A pedestrian's view on interacting particle systems, KPZ universality and random matrices {\it J. Phys. A: Math. Theor.} {\bf 43} 403001 
	
	\bibitem{Lazarescu2015} Lazarescu~A 2015 The physicist's companion to current fluctuations: one-dimensional bulk-driven lattice gases {\it J. Phys. A: Math. Theor.} {\bf 48} 503001
	
	\bibitem{Krug1991} Krug~J 1991 Boundary-induced phase transitions in driven diffusive systems {\it Phys. Rev. Lett.} {\bf 67} 1882

	\bibitem{DerridaDomanyMukamel1992} Derrida~B, Domany~E and Mukamel~D 1992 An exact solution of a one-dimensional asymmetric exclusion model with open boundaries {\it J. Stat. Phys.} {\bf 69} 667

	\bibitem{SchutzDomany1993} Sch\"utz~G and Domany~E 1993 Phase transitions in an exactly soluble one-dimensional exclusion process {\it J. Stat. Phys.} {\bf 72} 277

	\bibitem{DerridaEvansHakimPasquier1993} Derrida~B, Evans~M~R, Hakim~V and Pasquier~V 1993 Exact solution of a 1D asymmetric exclusion model using a matrix formulation {\it J. Phys. A: Math. Gen.} {\bf 26} 1493

	\bibitem{Derrida1998} Derrida~B 1998 An exactly soluble non-equilibrium system: the asymmetric simple exclusion process {\it Phys. Rep.} {\bf 301} 65

	\bibitem{JanowskyLebowitz1992} Janowsky~S~A and Lebowitz~J~L 1992 Finite-size effects and shock fluctuations in the asymmetric simple-exclusion process {\it Phys. Rev. A} {\bf 45} 618

	\bibitem{JanowskyLebowitz1994} Janowsky~S~A and Lebowitz~J~L 1994 Exact results for the asymmetric simple exclusion process with a blockage {\it J. Stat. Phys.} {\bf 77} 35

	\bibitem{Kolomeisky1998} Kolomeisky~A~B 1998 Asymmetric simple exclusion model with local inhomogeneity {\it J. Phys. A: Math. Gen.} {\bf 31} 1153 
	
	\bibitem{TripathyBarma1998} Tripathy~G and Barma~M 1998 Driven lattice gases with quenched disorder: Exact results and different macroscopic regimes {\it Phys. Rev. E} {\bf 58} 1911
	
	\bibitem{BengrineBenyoussefEzZahraouyMhirech1999} Bengrine~M, Benyoussef~A, Ez-Zahraouy~H and Mhirech~F 1999 Traffic model with quenched disorder {\it Phys. Lett. A} {\bf 253} 135
	
	\bibitem{StinchcombeDeQueiroz2011} Stinchcombe~R~B and de~Queiroz~S~L~A 2011 Smoothly varying hopping rates in driven flow with exclusion {\it Phys. Rev. E} {\bf 83} 061113

	\bibitem{BanerjeeBasu2020} Banerjee~T and Basu~A 2020 Smooth or shock: Universality in closed inhomogeneous driven single file motions {\it Phys. Rev. Res.} {\bf 2} 013025
	
	\bibitem{GoswamiChatterjeeMukherjee2022} Goswami~A, Chatterjee~M and Mukherjee~S 2022 Steady states and phase transitions in heterogeneous asymmetric exclusion processes {\it J. Stat. Mech.: Theor. Exp.} 123209

	\bibitem{PopkovRakosWillmannKolomeiskySchutz2003} Popkov~V, R\'akos~A, Willmann~R~D, Kolomeisky~A~B and Sch\"utz~G~M 2003 Localization of shocks in driven diffusive systems without particle number conservation {\it Phys. Rev. E} {\bf 67} 066117

	\bibitem{EvansJuhaszSanten2003} Evans~M~R, Juh\'asz~R and Santen~L 2003 Shock formation in an exclusion process with creation and annihilation {\it Phys. Rev. E} {\bf 68} 026117

	\bibitem{ParmeggianiFranoschFrey2003} Parmeggiani~A, Franosch~T and Frey~E 2003 Phase coexistence in driven one-dimensional transport {\it Phys. Rev. Lett.} {\bf 90} 086601

	\bibitem{ParmeggianiFranoschFrey2004} Parmeggiani~A, Franosch~T and Frey~E 2004 Totally asymmetric simple exclusion process with Langmuir kinetics {\it Phys. Rev. E} {\bf 70} 046101

	\bibitem{BottoPelizzolaPrettiZamparo2019} Botto~D, Pelizzola~A, Pretti~M and Zamparo~M 2019 Dynamical transition in the TASEP with Langmuir kinetics: mean-field theory {\it J. Phys. A: Math. Theor.} {\bf 52} 045001

	\bibitem{BottoPelizzolaPrettiZamparo2020} Botto~D, Pelizzola~A, Pretti~M and Zamparo~M 2020 Unbalanced Langmuir kinetics affects TASEP dynamical transitions: mean-field theory {\it J. Phys. A: Math. Theor.} {\bf 53} 345001

	\bibitem{KatzLebowitzSpohn1984} Katz~S, Lebowitz~J~L and Spohn~H 1984 Nonequilibrium steady states of stochastic lattice gas models of fast ionic conductors {\it J. Stat. Phys.} {\bf 34} 497
	
	\bibitem{PopkovSchutz1999} Popkov~V and Sch\"utz~G~M 1999 Steady-state selection in driven diffusive systems with open boundaries {\it Europhys. Lett.} {\bf 48} 257
	
	\bibitem{HagerKrugPopkovSchutz2001} Hager~J~S, Krug~J, Popkov~V and Sch\"utz~G~M 2001 Minimal current phase and universal boundary layers in driven diffusive systems {\it Phys. Rev. E} {\bf 63} 056110
	
    \bibitem{DierlMaassEinax2011} Dierl~M, Maass~M and Einax~M 2011 Time-dependent density functional theory for driven lattice gas systems with interactions {\it EPL} {\bf 93} 50003
    
    \bibitem{DierlMaassEinax2012} Dierl~M, Maass~M and Einax~M 2012 Classical driven transport in open systems with particle interactions and general couplings to reservoirs {\it Phys. Rev. Lett.} {\bf 108} 060603
    
    \bibitem{DierlEinaxMaass2013} Dierl~M, Einax~M and Maass~M 2013 One-dimensional transport of interacting particles: Currents, density profiles, phase diagrams, and symmetries {\it Phys. Rev. E} {\bf 87} 062126
    
	\bibitem{MidhaKolomeiskyGupta2018jstat} Midha~T, Kolomeisky~A~B and Gupta~A~K 2018 Effect of interactions for one-dimensional asymmetric exclusion processes under periodic and bath-adapted coupling environment {\it J. Stat. Mech.: Theor. Exp.} 043205

	\bibitem{PelizzolaPrettiPuccioni2019} Pelizzola~A, Pretti~M and~Puccioni F 2019 Dynamical transitions in a one-dimensional Katz-Lebowitz-Spohn model {\it Entropy} {\bf 21} 1028

	\bibitem{AntalSchutz2000} Antal~T and Sch\"utz~G~M 2000 Asymmetric exclusion process with next-nearest-neighbor interaction: Some comments on traffic flow and a nonequilibrium reentrance transition {\it Phys. Rev. E} {\bf 62} 83
	
	\bibitem{BottoPelizzolaPretti2018} Botto~D, Pelizzola~A and Pretti~M 2018 Dynamical transitions in a driven diffusive model with interactions {\it EPL} {\bf 124} 50004
    
	\bibitem{BaumgaertnerNarasimhan2023} Baumg\"artner~A and Narasimhan~S~L 2023 Phase transitions in the driven lattice gas (TASEP) with repulsive energies {\it J. Phys. A: Math. Theor.} {\bf 56} 355001
	
	\bibitem{PierobonMobiliaKouyosFrey2006} Pierobon~P, Mobilia~M, Kouyos~R and Frey~E 2006 Bottleneck-induced transitions in a minimal model for intracellular transport {\it Phys. Rev. E} {\bf 74} 031906

	\bibitem{MidhaKolomeiskyGupta2018pre} Midha~T, Kolomeisky~A~B and Gupta~A~K 2018 Interactions in nonconserving driven diffusive systems {\it Phys. Rev. E} {\bf 98} 042119
	
	\bibitem{JindalMidhaGupta2020} Jindal~A, Midha~T and Gupta~A~K 2020 Analysis of interactions in totally asymmetric exclusion process with site-dependent hopping rates: theory and simulations {\it J. Phys. A: Math. Theor.} {\bf 53} 235001
	
	\bibitem{PalGupta2021} Pal~B and Gupta~A~K 2021 Role of interactions in a closed quenched driven diffusive system {\it J. Phys. A: Math. Theor.} {\bf 54} 025005
	
	\bibitem{Pelizzola2005} Pelizzola~A 2005 Cluster variation method in statistical physics and probabilistic graphical models {\it J. Phys. A: Math. Gen.} {\bf 38} R309 

	\bibitem{PlischkeBergersen1994} Plischke~M and Bergersen~B 1994 {\it Equilibrium statistical physics} (Singapore: World Scientific) 

	\bibitem{PelizzolaPretti2017} Pelizzola~A and Pretti~M 2017 Cluster approximations for the TASEP: stationary state and dynamical transition {\it Eur. Phys. J. B} {\bf 90} 183

	\bibitem{KolomeiskySchutzKolomeiskyStraley1998} Kolomeisky~A~B, Sch\"utz~G~M, Kolomeisky~E~B and
	Straley~J~P 1998 Phase diagram of one-dimensional driven lattice gases with
	open boundaries {\it J. Phys. A: Math. Gen.} {\bf 31} 6911

	\bibitem{ReichenbachFranoschFrey2008} Reichenbach~T, Franosch~T and Frey~E 2008 Domain wall delocalization, dynamics and fluctuations in an exclusion process with two internal states {\it Eur. Phys. J. E} {\bf 27} 47 

	\bibitem{MironReuveni2021} Miron~A and Reuveni~S 2021 Diffusion with local resetting and exclusion {\it Phys. Rev. Res.} {\bf 3} L012023

	\bibitem{PelizzolaPrettiZamparo2021} Pelizzola~A, Pretti~M and Zamparo~M 2021 Simple exclusion processes with local resetting {\it EPL} {\bf 133} 60003

	\bibitem{PelizzolaPretti2022} Pelizzola~A and Pretti~M 2022 Totally asymmetric simple exclusion process with local resetting and open boundary conditions {\it J. Phys. A: Math. Theor.} {\bf 55} 454001

	\bibitem{NagarGupta2023} Nagar~A and Gupta~S 2023 Stochastic resetting in interacting particle systems: a review {\it J. Phys. A: Math. Theor.} {\bf 56} 283001


%
%
%
%
%
%
%
%
%
%
%
%
%
%
%
%
%
%
%
%
	
\end{thebibliography}
\end{document}